\documentclass[aps,prx,reprint,superscriptaddress]{revtex4-2}
\usepackage[T1]{fontenc}
\usepackage[utf8]{inputenc}
\usepackage{color}
\usepackage{array}
\usepackage{float}
\usepackage{multirow}
\usepackage{amsmath}
\usepackage{amsthm}
\usepackage{amssymb}
\usepackage{graphicx}
\usepackage{comment}

\makeatletter

\providecommand{\tabularnewline}{\\}

\usepackage{hyperref}
\usepackage{siunitx}

\makeatother

\usepackage{babel}
\begin{document}
\title{Shadow tomography for classical tensor network simulations}

\author{Jiace Sun}
\author{Garnet Kin-Lic Chan}
\email{gkc1000@gmail.com}
\affiliation{Marcus Center for Theoretical Chemistry, California Institute of Technology, Pasadena, CA 91125, USA}
\affiliation{Division of Chemistry and Chemical Engineering, California Institute of Technology, Pasadena, CA 91125, USA}

\begin{abstract}
Shadow tomography has appeared as a powerful tool for estimating observables on quantum computers from a small number of samples. We show that shadow-tomography-inspired ideas can offer similarly improved sample scaling for estimating observables on tensor network states on classical computers after proper adaptation.
We develop strategies for both spin (bosonic) and fermionic systems, tailored to the contraction requirements of tensor networks, and generate scaling improvements of factors of $O(N)$ to $O(N^{3})$ (where $N$ is system size), depending on the specific task and system type.
For the important and difficult task of evaluating the expectation value of long-range interacting Hamiltonians, we achieve the optimal $O(1)$ overall scaling (up to logarithmic factors) for an arbitrarily fixed relative Monte Carlo error in both spin and fermionic systems.
Additionally, we show that shadow estimators offer more stable gradients of observables in variational optimization tasks than standard Monte Carlo estimators. 
We demonstrate practical advantage by simulating systems with long-range interactions, including the 2D long-range Heisenberg model and an ab-initio quantum chemistry Hamiltonian.
\end{abstract}
\maketitle

\section{Introduction}

Tensor network (TN) methods have become one of the central tools for the study of quantum many-body systems on classical computers.
\cite{RevModPhys.93.045003,ORUS2014117,RevModPhys.77.259,SCHOLLWOCK201196,Verstraete01032008,orus2019tensor} By exploiting the structure of physically relevant wavefunctions, they provide efficient representations of states whose full Hilbert-space dimension grows exponentially with system size.
Tensor networks have therefore played a major role in the simulation of strongly correlated lattice models \cite{liu2025accurate,PhysRevLett.101.090603,PhysRevB.81.174411}, quantum dynamics \cite{PhysRevLett.93.076401,vidal2004efficient}, and electronic structure \cite{annurev:/content/journals/10.1146/annurev-physchem-032210-103338,10.1063/1.5129672} problems.
In practice, however, representing the state is only part of the computational task.
A comparably important challenge is the evaluation of physical observables from the represented state, such as correlation functions, reduced density matrices (RDMs), and Hamiltonian expectation values.

For simple local Hamiltonians and low-dimensional tensor networks, many observables can be evaluated efficiently.
More generally, however, observable evaluation itself can become a major bottleneck when the number of relevant quantities is large.
This occurs, for example, when one seeks the full set of spatial correlation functions in a many-body system \cite{PhysRevX.8.041033}, and when evaluating long-range model Hamiltonians \cite{zhao2023finite} containing many two-body terms, or in \textit{ab initio} quantum chemistry \cite{keller2015efficient}, where physical quantities are naturally expressed through complete one- and two-body RDMs.
Existing tensor network approaches mainly rely either on direct contraction of the corresponding expectation-value networks \cite{nishino1996corner,PhysRevLett.101.250602,lubasch2014unifying,PhysRevLett.132.117401,ran2020tensor}, or on Monte Carlo sampling from the Born distribution \cite{PhysRevLett.99.220602,PhysRevB.83.134421,PhysRevB.95.195154,liu2025accurate}.
Although these two strategies are algorithmically quite different, both remain strongly affected by the total number of observables or Hamiltonian terms: direct contraction must evaluate many expectation values separately (or in the case of the Hamiltonian, contract a complicated operator), while Monte Carlo methods still require separate amplitude evaluations for general non-diagonal operator terms.

This bottleneck suggests a different question: can one design an estimator in which a single sample simultaneously contains information about many observables?
On quantum computers, a powerful affirmative answer is provided by shadow tomography \cite{aaronson2020shadow,aaronson2019gentle,PhysRevA.107.042403}, which targets a more restricted task than state tomography: rather than learning the state $\rho$ in full, it estimates only a specified collection of $M$ observable expectation values $\{\mathrm{Tr}(O_{i}\rho)\}$, with $M$ typically far smaller than the number of parameters needed to specify $\rho$.
A well-known example is the \emph{classical shadow} method of Huang, Kueng, and Preskill \cite{huang2020predicting}, which uses randomized measurements \cite{elben2023randomized} together with an inverse measurement channel to construct an unbiased state estimator $\hat{\rho}$ satisfying $\mathbb{E}[\hat{\rho}]=\rho$.
Many variants have since been developed, including shallow Clifford shadows \cite{bertoni2024shallow}, fermionic and matchgate shadows \cite{PhysRevLett.127.110504,wan2023matchgate,heyraud2025unified}, locally-biased shadows \cite{hadfield2022measurements}, derandomized shadows \cite{huang2021efficient}, shadows based on locally-scrambled dynamics \cite{hu2023classical}, and tensor network assisted classical shadows \cite{Akhtar2023scalableflexible}.
More generally, the shadow-tomography umbrella also covers other structured measurement schemes such as commuting-family measurements \cite{verteletskyi2020measurement,yen2020measuring,huggins2021efficient} and Bell-basis measurements \cite{PhysRevLett.133.020601,PRXQuantum.6.010336,fq8z-y55j} that share the same one-sample-many-observables structure.
In all cases, the measurement complexity of estimating many observables can be exponentially or polynomially improved from the naive scaling proportional to the number of observables.

While shadow tomography was originally developed for quantum computers, the underlying idea is more general: structured measurements can be viewed as a way of compressing information about many observables into a single sample.
This naturally raises the possibility that similar ideas may also be useful for classical simulations.
However, the transfer of these ideas has been limited: existing applications of shadow tomography to classical simulation include an estimation of entanglement measures in tensor networks (which suffers from sampling complexity exponential in subsystem size \cite{feldman2022entanglement}), or the use of Bell sampling in the context of the stochastic series expansion
\cite{fq8z-y55j}.

In this work, we show that several shadow-tomography ideas can be adapted to tensor network simulations on classical computers in the context of variational Monte Carlo simulation \cite{becca2017quantum}, leading to substantially improved asymptotic scaling for observable evaluation.
To make the comparison precise, we consider two tasks: estimating all elements of a $k$-RDM to fixed additive accuracy, and estimating a scalar physical observable that is a linear function of the $k$-RDM, such as a long-range spin Hamiltonian or an \textit{ab initio} electronic Hamiltonian, to fixed relative accuracy.
For spin (bosonic) tensor network states, we develop a Pauli-shadow-type strategy that compresses information on multiple local observables into each Monte Carlo sample, achieving $O(N)$ improvement for both tasks in the physically most relevant case of $k=2$.
For fermionic tensor network states, where naive Pauli shadows become inefficient because of the nonlocality induced by Jordan-Wigner strings, we instead develop alternative strategies based on structured commuting measurements and Bell-basis measurements, which we call the \emph{rainbow-basis shadow} and the \emph{Bell-sampling shadow} methods respectively, resulting in an $O(N)$ improvement for estimating the $2$-RDM, and an $O(N^{3})$ improvement when estimating fermionic $2$-body physical observables.
With the standard environment reuse strategy, for both spin and fermionic systems, shadow estimators achieve the optimal $O(1)$ overall sample scaling for evaluating the Hamiltonian expectation values to an arbitrarily fixed Monte Carlo relative error.
We further show that these shadow estimators provide a bounded-variance gradient estimator for observables with respect to variational parameters in the state, leading to improved stability in tensor-network optimization tasks.
We verify the predicted theoretical scalings and numerical stabilities in numerical calculations on long-range Heisenberg models and ab initio hydrogen chains, and further show that a practical advantage can already be reached for variational tensor network ground-state simulations over existing Monte Carlo methods.

Our results show that shadow estimators can serve as a useful new algorithmic tool for classical tensor network simulations.
They provide a way to reduce the observable-evaluation bottleneck in settings where the number of relevant observables is large, while remaining compatible with the tensor network structures that make many-body simulation feasible.
More broadly, this work illustrates that ideas developed for quantum computation can improve classical many-body algorithms as well, opening new connections between quantum information and tensor network simulation.

\section{Theory}

\subsection{Reduced density matrices and evaluation of observables}
A central task in the study of quantum many-body systems is the evaluation of expectation values of observables supported on a small subset of degrees of freedom.
For a pure quantum state $|\psi\rangle$ defined on a lattice of $N$ sites, the properties of any subset $S$ of sites are completely characterized by the reduced density matrix of that subsystem, $\rho_{S}=\mathrm{Tr}_{\bar{S}}\left(|\psi\rangle\langle\psi|\right),$where $\bar{S}$ denotes the complement of $S$.
Consequently, the reduced density matrix of $k$ sites contains all information required to evaluate arbitrary $k$-local observables.
This connection implies that $k$-site reduced density matrices are equivalent to collections of $k$-point correlation functions.
For example, in spin systems the two-site reduced density matrix contains the same information as the set of all two-point correlations $\langle\sigma_{i}^{\alpha}\sigma_{j}^{\beta}\rangle,$where $\sigma_{i}^{\alpha}$ denotes a Pauli operator acting on site $i$, and $\alpha=x,y,z$.
Similarly, in fermionic systems the two-body reduced density matrix contains expectation values of operators of the form $\langle c_{i}^{\dagger}c_{j}^{\dagger}c_{k}c_{l}\rangle$ (with some other similar blocks for systems without $U(1)$ symmetry) and is sufficient to compute the expectation values of Hamiltonians with up to two-body interactions.
While some simple models require only a small subset of two-point or two-body correlations, more general long-range or ab-initio Hamiltonians can involve all such interaction terms and therefore depend on the complete RDMs (site-based for spin systems and particle-based for fermions).

Tensor network states provide an efficient representation of many-body wavefunctions whose full Hilbert-space description would otherwise be exponentially large.
In tensor network simulations, expectation values of observables or the elements of reduced density matrices are typically obtained by contracting the tensor network $\langle\psi_{TN}|O|\psi_{TN}\rangle$.
Existing tensor-network approaches for computing these quantities broadly fall into two paradigms.
The first evaluates expectation values by contracting the corresponding double-layer tensor network, either exactly in one-dimensional systems (i.e. matrix product states (MPS)) \cite{ORUS2014117}, or through controlled approximation in higher dimensions (i.e. projected entangled pair states (PEPS)).
\cite{nishino1996corner,PhysRevLett.101.250602,PhysRevLett.132.117401} This clearly results in a computational scaling proportional to the number of local observables $O$.
The second employs Monte Carlo samples \cite{PhysRevX.8.041033,PhysRevB.83.134421}, in which computational-basis configurations $|x\rangle$ are sampled from the Born probability, $p(x)=|\langle x|\psi\rangle|^{2},$ and then used to evaluate the expectation values as
\begin{equation}
\begin{aligned}\langle O\rangle & =\mathbb{E}_{x\sim p(\cdot)}\left[\frac{\langle x|O|\psi\rangle}{\langle x|\psi\rangle}\right]\\
O_{\text{loc}}^{\text{standard}}(x) & =\frac{\langle x|O|\psi\rangle}{\langle x|\psi\rangle}
\end{aligned}
\label{eq:standard_MC}
\end{equation}
where in the general case $O|x\rangle$ yields another configuration $|x'\rangle$, or a linear combination of configurations.
The sampling of configurations $|x\rangle$ can be achieved by Markov Chain Monte Carlo (MCMC), thus such an approach requires only single-layer contractions, with the price being the need to use a large number of samples, as controlled by the sampling variance and required accuracy.
Again, the single-layer contraction is exact for 1D MPS and approximate for higher dimensions.
For different local observables $O$, $\langle x|O|\psi\rangle$ generally involves different amplitudes $|x'\rangle$.
Therefore, the computational cost is again approximately proportional to the number of local observables.

To make a fair comparison between different approaches, we consider two tasks in this work: (1) estimating all the $k$-RDM elements to additive error $\epsilon$ (with a high probability), and (2) estimating some given scalar physical observable (such as the Hamiltonian) expressed as a linear function of the $k$-RDM, to multiplicative error $\epsilon_{r}\sim\epsilon/N$.
For task (2), focusing on $k=2$, the general forms are
\begin{equation}
H_{\text{spin}}=-\sum_{i\alpha}h_{i\alpha}\sigma_{i}^{\alpha}-\sum_{ij\alpha\beta}J_{ij}^{\alpha\beta}\sigma_{i}^{\alpha}\sigma_{j}^{\beta}\label{eq:H_spin}
\end{equation}
for long-range spin Hamiltonians, and
\begin{equation}
H_{\text{fermion}}=\sum_{pq}h_{pq}p^{\dagger}q+\sum_{pqrs}V_{pqrs}p^{\dagger}q^{\dagger}sr\label{eq:H_fermion}
\end{equation}
for ab initio fermionic quantum chemistry Hamiltonians.
We additionally comment that, another approach to perform task (2) is via contraction with a single tensor network operator representation of the observable, which superficially does not require one to estimate each component of the operator individually. However, while we will not consider these approaches in detail further, the bond dimension of the tensor network operator still grows with the number of terms in the sums in Eqs.~\ref{eq:H_spin}, ~\ref{eq:H_fermion}, and thus the cost again depends on the number of involved terms in the $k$-RDM (and may in fact grow superlinearly with this number via the dependence on the bond dimension)~\cite{chan2016matrix,o2020simplified}.

We first give the ``baseline'' scaling of these two tasks for the two standard approaches above.
Let $M_{k}$ be the number of $k$-RDM elements, i.e. $M_{k}\sim O(N^{k})$ for spin systems and $M_{k}\sim O(N^{2k})$ for fermionic systems.
For the direct contraction strategy, the two tasks are essentially equivalent, thus $M_{k}$ double-layer contractions are needed.
For the standard MC strategy, the variance of each RDM element is $O(1)$, thus the scaling is $O(M_{k}/\epsilon^{2})$ for task (1).
If $\epsilon$ is the maximum error instead of average error, an extra $\log(M_{k})\sim\log N$ factor is needed to account for the exponential tail of the Gaussian distribution.
For task (2), we can generally express the final scaling as
\begin{equation}
\begin{alignedat}{1}\text{Total cost} & =(\text{Variance of evaluation})\\
 & \times(\text{TN contractions per evaluation)}\\
 & \times\text{(TN contraction cost)}\times\frac{1}{\epsilon^{2}}
\end{alignedat}
\label{eq:obs_var}
\end{equation}
where ``TN contractions per evaluation'' also includes the number of TN contractions to decorrelate the MCMC samples.
The standard MC approach requires $O(M_{k})$ TN contractions per evaluation, and the evaluation variance is $\langle(O-\langle O\rangle)^{2}\rangle$ (where $\langle\cdot\rangle=\langle\psi|\cdot|\psi\rangle$).
For physical size-extensive observables, the variance has a $O(N^{2})$ upper bound.
However, if we additionally assume that the quantum states have rapidly decaying correlations, the fluctuations from distant parts are decoupled, which leads to a reduced $O(N)$ variance.
Therefore, $O(NM_{k}/\epsilon^{2})\sim O(N^{-1}M_{k}/\epsilon_{r}^{2})$ single-layer TN contractions are needed.
These baseline results are summarized in Table~\ref{table:scaling}.

\subsection{Shadow estimators for tensor network simulations} \label{subsec:shadow}
We use the term \emph{shadow estimator} in the broad shadow-tomography sense: a sampling procedure whose outcomes provide simultaneous estimators for many observables, while not necessarily reconstructing the full state.
In the following, we introduce three strategies for classical tensor network simulation, realized through three different measurement schemes---randomized local-Clifford measurements, commuting-family measurements, and two-replica Bell measurements---which we refer to as the Pauli shadow, the rainbow-basis shadow, and the Bell-sampling shadow methods respectively.
Of these, only the Pauli construction is a strict classical shadow in the sense of Huang, Kueng, and Preskill \cite{huang2020predicting}, with an unbiased state estimator $\hat{\rho}$ satisfying $\mathbb{E}[\hat{\rho}]=\rho$; the other two are shadow-tomography-inspired estimators built directly from the relevant commuting structure.
All three strategies achieve reduced asymptotic scaling compared with the two standard approaches above (exact double-layer contraction, and Monte Carlo sampling).
A summary of the comparison between the shadow-estimator approaches and standard approaches is listed in Table~\ref{table:scaling}.

\begin{table*}
\caption{Computational scaling of two observable evaluation tasks for qubit/spin (bosonic) and fermionic systems on classical tensor network states using the approaches considered in this work. 
The two tasks are (1) evaluating all elements of a $k$-RDM to additive error $\epsilon$, and (2) evaluating the expectation value of a 2-body long range Hamiltonian (such as the long-range Heisenberg model for spins, and the ab-initio quantum chemistry Hamiltonian for fermions) to relative (multiplicative) error $\epsilon$. 
The shadow estimators include the standard Pauli shadow and the two shadow-tomography-inspired estimators based on commuting measurements (rainbow-basis shadow) and Bell-basis measurements (Bell-sampling shadow). 
$C_{1}$ and $C_{2}$ represent the cost of single-layer contraction and double-layer contraction, respectively. 
For a tensor network with $N$ sites and bond dimension $D$, typically $C_{i=1,2}=O(ND^{\alpha_{i}})$, with $\alpha_{2}>\alpha_{1}$, where the values of $\alpha_{1,2}$ depend on the tensor network geometry and the contraction algorithm. 
All $O(\log N)$ (or $O(\text{polylog}(N))$) factors are neglected. 
For the task of the Hamiltonian expectation value, we additionally assume that the Hamiltonian has sufficiently rapidly decaying interactions and that the quantum states have fast decaying correlations.
All scalings are obtained without the environment reuse strategy. 
If environments are reused, all listed scalings can be further reduced by $O(N)$, thereby achieving an overall $O(1)$ scaling for arbitrarily fixed relative error for both spin and fermionic 2-body Hamiltonians using the Pauli shadow and Bell sampling shadow estimators, respectively.
Bold indicates the best results among the compared strategies.}
\label{table:scaling}

\renewcommand{\arraystretch}{1.5}

\begin{tabular}{|c|c|c|c|c|c|}
\hline 
\multirow{3}{*}{System} & \multirow{3}{*}{Method} & \multicolumn{3}{c|}{Task} & \multirow{3}{*}{Unit}\tabularnewline
\cline{3-5}
 &  & \multirow{2}{*}{$k$-RDM} & \multicolumn{2}{c|}{2-body Hamiltonian} & \tabularnewline
\cline{4-5}
 &  &  & Theoretical & Numerical & \tabularnewline
\hline 
\multirow{3}{*}{Bosonic (spin)} & Direct contraction & $O(N^{k})$ & \multicolumn{2}{c|}{$O(N^{2})$} & $C_{2}$\tabularnewline
\cline{2-6}
 & Standard MC & $O(N^{k})$ & $O(N)$ & $O(N^{1.1})$ & $C_{1}/\epsilon^{2}$\tabularnewline
\cline{2-6}
 & Pauli shadow & \textbf{$\boldsymbol{O(N)}$} & $\boldsymbol{O(1)}$ & $\boldsymbol{O(N^{0.1})}$ & $C_{1}/\epsilon^{2}$\tabularnewline
\hline 
\multirow{4}{*}{Fermionic} & Direct contraction & $O(N^{2k})$ & \multicolumn{2}{c|}{$O(N^{4})$} & $C_{2}$\tabularnewline
\cline{2-6}
 & Standard MC & $O(N^{2k})$ & $O(N^{3})$ & $O(N^{3.4})$ & $C_{1}/\epsilon^{2}$\tabularnewline
\cline{2-6}
 & Rainbow shadow & $\boldsymbol{O(N^{k+1})}$ & $O(N^{2})$ & N/A & $C_{1}/\epsilon^{2}$\tabularnewline
\cline{2-6}
 & Bell sampling & $^{\star}\Omega(N^{k+1})$ & $\boldsymbol{O(1)}$ & \textbf{$\boldsymbol{O(N^{0.2})}$} & $C_{1}/\epsilon^{2}$\tabularnewline
\hline 
\end{tabular}

$^{\star}$: A slightly different construction in ~\cite{PRXQuantum.6.010336} can give $O(C_2 N \log N /\epsilon^4)$ complexity, but requires double-layer contraction.

\end{table*}


\subsubsection{Qubit/spin (bosonic) k-RDM by Pauli shadow} \label{subsec:pauli}
We first consider evaluating the $k$-RDM (assume $k$ is $O(1)$) for qubit/spin systems.
On quantum computers, this can be efficiently computed by the Pauli shadow method, which involves two steps: (1) choose a unitary from some pre-defined ensemble and apply it to the state, and (2) measure in the computational basis.
The unitary ensemble is the product of single-qubit uniformly random Clifford or Haar-random unitaries.
Let the measurement result be $|x\rangle=|x_{1}x_{2}...x_{N}\rangle$, the selected unitary be $U=U_{1}\otimes U_{2}\otimes...\otimes U_{N}$, then the unbiased estimator is
\begin{equation}
\rho_{s}(U,x)=\otimes_{i=1}^{N}(3U_{i}^{\dagger}|x_{i}\rangle\langle x_{i}|U_{i}-I_{2}) \label{eq:pauli_shadow}
\end{equation}
where $I_{2}$ is the 2$\times$2 identity operator.
Since $\rho_{s}(U,x)$ is a product state, the expectation value of any Pauli operator $P$ can be efficiently evaluated classically.

For an arbitrary $k$-local Pauli operator, this unbiased estimator gives an $O(3^{k})$ variance.
Thus for small and fixed $k$, one measurement gives $O(1)$ variance for all $k$-local Pauli operators, or equivalently the complete $k$-RDM.

To extend it to classical tensor network simulations, we need to perform two operations classically: (1) apply the corresponding unitary operation, and (2) perform the ``quantum measurement''.
The former one is simple in this Pauli shadow case, as the single-qubit unitary can be locally applied to a tensor network.
To classically perform the ``quantum measurement'', we use the Metropolis-Hastings algorithm.
Naively, one needs to sample both $U$ and $x$ satisfying the required distribution $p(U,x;\rho)=p(U)\langle x|U\rho U^{\dagger}|x\rangle$.
However, since $p(U)$ is a uniform distribution for both the Clifford and Haar-random case, one can observe that both the sampling probability $p(U,x;\rho)$ and the constructed unbiased sample $\rho_{s}(U,x)$ depend only on $U^{\dagger}|x\rangle$.
Thus, the process can be simplified to sampling only $U^{\dagger}|x\rangle$, which is a product pure state, and which we denote  as $|\phi\rangle=\otimes_{i=1}^{N}|\phi_{i}\rangle$.
Therefore, we apply MCMC by sequentially updating each single-qubit state $|\phi_{i}\rangle$ to either a uniformly random stabilizer state (i.e. the set of states closed under Clifford operations), or a Haar-random state, then accept with the probability $\min(p_{\text{new}}/p_{\text{old}},1)$, where $p_{I}=|\langle\phi_{I}|\psi_{TN}\rangle|^{2}$, $I=\text{new},\text{old}$.
For states with area-law entanglement and fast decaying correlations, we expect local Metropolis updates to have an $O(1)$ acceptance rate, thus the Markov chain mixing time is expected to be $O(N)$ (as one can touch all sites with $O(N)$ updates).
Therefore, the full $k$-RDM can be evaluated to additive error $\epsilon$ with $O(N/\epsilon^{2})$ number of single-layer contractions (note we do not consider any tensor network reuse until Sec.~\ref{subsec:reuse}).

Finally, we provide an estimate of the variance of scalar physical observables expressed as linear functions of the $k$-RDM for $k=1,2$ (the example of $k=2$ is shown in Eq.
\ref{eq:H_spin}). The sampled value of each Pauli operator is $O(1)$ (its absolute value is bounded by $3^{k}$). With additional fast decay of interactions as in many physical problems, 
the sampled total expectation value has a $O(N)$ upper bound, leading to an $O(N^{2})$ variance upper bound.
If the quantum state is assumed to have rapidly decaying correlations, the fluctuations on distant  sites are decoupled, therefore, we expect the variance to be $O(N)$.
Combined with the $O(N)$ Markov chain mixing time, we expect $O(N^{2}/\epsilon^{2})\sim O(1/\epsilon_{r}^{2})$ single-layer contractions are needed to evaluate the expectation value to additive error $\epsilon$, or relative error $\epsilon_{r}\sim\epsilon/N$.

\subsubsection{Fermionic $k$-RDM by rainbow-basis shadow} \label{subsec:rainbow}
\begin{figure*}
\centering
\includegraphics[width=0.9\linewidth]{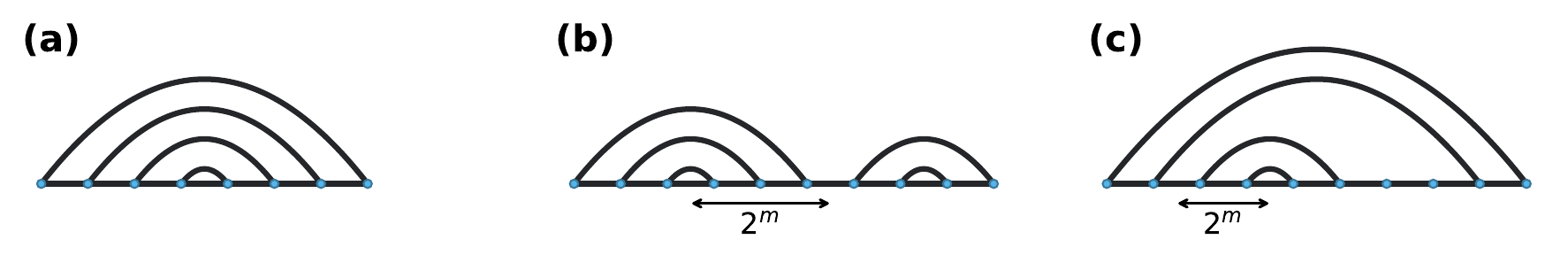}

\caption{Schematic illustration of the fermionic rainbow-basis shadow construction for MPS. (a): single rainbow Bell pairs for $1$-RDM measurements. (b, c): two forms of double-rainbow Bell pairs for 2-RDM measurements. In the 2-RDM case, the width of the rainbow is denoted $2^m$, for the detailed (nearly optimal) sampling scheme described in the Appendix. \label{fig:rainbow}}
\end{figure*}

The Pauli shadow does not naively work for the fermionic $k$-RDM, as the fermionic operators become non-local after Jordan-Wigner transformation, resulting in exponential variance (this can be reduced to a polynomial scaling using the Bravyi-Kitaev encoding~\cite{bravyi2002fermionic} or an advanced encoding based on ternary trees~\cite{jiang2020optimal}, but such transformations break geometric locality that is vital for tensor network states).
A few approaches have been specifically designed to evaluate fermionic $k$-RDM on (non-fermionic) quantum computers by accessing Haar-random fermionic Gaussian operations \cite{wan2023matchgate,heyraud2025unified} or random qubit reordering followed by Pauli shadows \cite{PhysRevLett.127.110504}.
Both strategies achieve the asymptotically optimal $O(N^{k}/\epsilon^{2})$ scaling of number of measurements for all $k$-RDM elements with additive error $\epsilon$ (this number comes from the maximum number of commuting $k$-RDM elements in any basis, which determines the maximum amount of information that can be extracted from a single measurement) \cite{PRXQuantum.6.010336}.
However, both of these methods are infeasible for classical (fermionic) tensor network states, as random fermionic Gaussian operations and reordering are inefficient to implement in  tensor network algorithms.
Here we present an MPS-specific approach with an asymptotically optimal $O(N/\epsilon^{2})$ samples for the $1$-RDM, and a near asymptotically optimal $O(N^{2}\log^{2}N/\epsilon^{2})$ number of samples for the $2$-RDM, while using only operations that are efficiently implementable for MPS.
The total sample scaling can be obtained by multiplying by the $O(N)$ Markov chain mixing time required for independent samples.

For fermionic $k$-RDMs, it is common to transform the fermionic annihilation and creation operators to the Majorana basis, defined as
\begin{equation} \label{eq:majorana}
\begin{aligned}\gamma_{2i-1} & =a_{i}+a_{i}^{\dagger}\\
\gamma_{2i} & =-i(a_{i}-a_{i}^{\dagger})
\end{aligned}
\end{equation}
The corresponding $k$-RDM operators in the Majorana basis are products of $2k$ Majorana operators.
For notational convenience, we also write the two Majoranas associated with site $i$ as $\gamma_{i,p}\equiv\gamma_{2i-p}$, with a binary flavor label $p\in\{0,1\}$.
A fermionic $k$-RDM element therefore expands into at most $2^{2k}$ fixed-flavor Majorana strings.
In the construction below, the grouping is applied separately to each fixed flavor pattern $(p_{1},\ldots,p_{2k})$.
Since $k$ is fixed in the applications considered here, this only changes the prefactor and does not affect the asymptotic scaling.
The key observation is that the $O(N^{k})$ scaling for the class of fermionic shadows comes from the fact that one can divide the $O(N^{2k})$ $2k$-Majorana strings into the $O(N^{k})$ mutually commuting subsets.
The use of commuting measurements to estimate many compatible observables from a single set of samples has a long history in the VQE / Hamiltonian-estimation literature \cite{verteletskyi2020measurement,yen2020measuring,huggins2021efficient}.
Let a set of mutually commuting operators be $O_{1},...,O_{M}$, which are simultaneously diagonalized in the basis $|\alpha\rangle$.
Then the expectation values of these operators can be simultaneously evaluated as
\begin{equation}
\text{Tr}[O_{i}\rho]=\mathbb{E}_{\alpha\sim\langle\cdot|\rho|\cdot\rangle}\langle\alpha|O_{i}|\alpha\rangle.\label{eq:commuting_evaluation}
\end{equation}
To sample $\alpha$ from probability $p(\alpha)=\langle\alpha|\rho|\alpha\rangle$, we notice that the basis has a $Z_{2}^{N}$ structure, which allows us to write $|\alpha\rangle=|\alpha_{1},...,\alpha_{N}\rangle$, where the subscripts represent some (not necessarily local) binary labeling.
For classical tensor network simulations, we can apply the Metropolis-Hastings algorithm to sequentially update $\alpha_{1},...,\alpha_{N}$.
However, two extra requirements are needed: (1) the labelling $\alpha_{i}$ should not encode fully global information, otherwise the acceptance rate can be exponentially small, and (2) the computation of $\langle\alpha|\rho|\alpha\rangle$ (or $\langle\alpha|\psi\rangle$ for pure states) must be classically efficient.

Now we present a subset division of the $O(N^{2k})$ $2k$-Majoranas for $k=1,2$ that satisfies the above two conditions.
For $k=1$, fix the two Majorana flavor labels $p,q\in\{0,1\}$ and consider operators $\gamma_{i,p}\gamma_{j,q}$.
For each fixed $(p,q)$, we construct subsets indexed by $r$ as
\begin{equation}
S_{r}^{pq}=\{\gamma_{i,p}\gamma_{j,q}\,|\,i+j=r\}.
\end{equation}
One can verify that all operators in $S_{r}^{pq}$ commute for any $r$.
Since $r$ has $O(N)$ possible values and there are only four choices of $(p,q)$, all $2$-Majorana operators are divided into $O(N)$ mutually commuting subsets.
To define the measurement basis, each measured Majorana pair $\gamma_{i,p}\gamma_{j,q}$ is completed by the complementary pair $\gamma_{i,1-p}\gamma_{j,1-q}$ on the same physical sites.
The common eigenstates are therefore rainbow-like two-site Bell-pair states, together with local occupation states on any unpaired sites.
For example, in the common cases shown in Fig.~\ref{fig:rainbow} (a), the state takes the form $\otimes_{i+j=r}|\text{bell}\rangle_{ij}$, where $|\text{bell}\rangle$ is one of $\frac{1}{\sqrt{2}}\left(|00\rangle+e^{i\phi}|11\rangle\right)$ and $\frac{1}{\sqrt{2}}\left(|01\rangle+e^{i\phi}|10\rangle\right)$ with $\phi=0,\frac{\pi}{2},\pi,\frac{3\pi}{2}$. \footnote{The $\otimes_{i+j=r}|\text{bell}\rangle_{ij}$ notation indicates that the fermionic operators are applied in pairs, i.e. $\otimes_{i+j=r}\frac{1}{\sqrt{2}}(1+e^{i\phi}a_{i}^{\dagger}a_{j}^{\dagger})|\text{vac}\rangle$ or $\otimes_{i+j=r}\frac{1}{\sqrt{2}}(a_{i}^{\dagger}+e^{i\phi}a_{j}^{\dagger})|\text{vac}\rangle$.}
Note that the standard Bell states include only $\phi=0,\pi$, and here we have added two imaginary phases $\frac{\pi}{2},\frac{3\pi}{2}$, which correspond to the common eigenstates of Pauli operators $XY$ and $YX$.
For a fermionic MPS (which is equivalent to an MPS with $Z_{2}$ symmetry), one can evaluate the overlap by contracting the MPS from the center at $r$ (see Fig.~\ref{fig:rainbow}).
Such contractions can be performed exactly with a $O(ND^{3})$ complexity (where $D$ is the MPS bond dimension).
The sampling of the Bell pair states can be performed by Metropolis sampling, and the new candidate at each step can be constructed by locally updating a single Bell pair state.
If $r$ is not at the exact center of the MPS, there will be some sites that are not connected by any rainbow lines. These remaining sites can be sampled in the standard bitstring basis by the usual Metropolis scheme.

Now we sketch how to extend the strategy to $k=2$.
For a fixed Majorana flavor pattern, the 2-RDM Majorana strings can be written as $\gamma_{a,p_a}\gamma_{b,p_b}\gamma_{c,p_c}\gamma_{d,p_d}$ with site indices $a<b<c<d$.
We consider two grouping strategies: grouping operators with the same $a+b$ and $c+d$; and operators with the same $a+d$ and $b+c$ (see Fig. \ref{fig:rainbow} (b) and (c)).
Since each group contains two ``rainbows'', we must ensure that the rainbows do not overlap, otherwise the overlapping operators will not commute. A simple counting shows then that we have $O(N^2)$ possible values for the 2 centers, and each rainbow can sample $O(N)$ widths, leading to an overall $O(N^3)$ sampling cost. However, in this naive scheme different choices of rainbow widths contain redundancies with each other in terms of the sets of commuting operators. A more careful algorithm, which we describe in the Appendix~\ref{app:rainbow_2rdm}, needs to only consider $m = O(\log N)$ widths, leading to an overall near-optimal $O((N\log N)^{2})$ scaling.
We also conjecture that this strategy can be extended to general $k$-RDMs with $O(N^{k}\text{polylog}(N))$ scaling (although few physical applications require $k>2$).

We call the above constructions the \emph{rainbow-basis shadow} because the simultaneous-eigenbasis used in the measurement step consists of long-range two-site Bell pairs arranged in a rainbow pattern, as shown in Fig.~\ref{fig:rainbow}.

With the assumption of fast decay of correlations similar to as discussed previously, we expect the variance scaling of physical extensive observables that are linear functions of the $k$-RDM to be $O(N)$.
Counting the $O(N^{k})$ samples (neglecting $O(\log N)$ factors) required for each complete $2$-RDM evaluation, and the $O(N)$ Markov chain mixing time, we expect $O(N^{4}/\epsilon^{2})\sim O(N^{2}/\epsilon_{r}^{2})$ single-layer contractions will be needed to reach fixed additive error $\epsilon$, or relative error $\epsilon_{r}$.

\subsubsection{Fermionic $k$-RDM by Bell-sampling shadow} \label{subsec:bell}
\begin{figure}
\centering
\includegraphics[width=0.9\columnwidth]{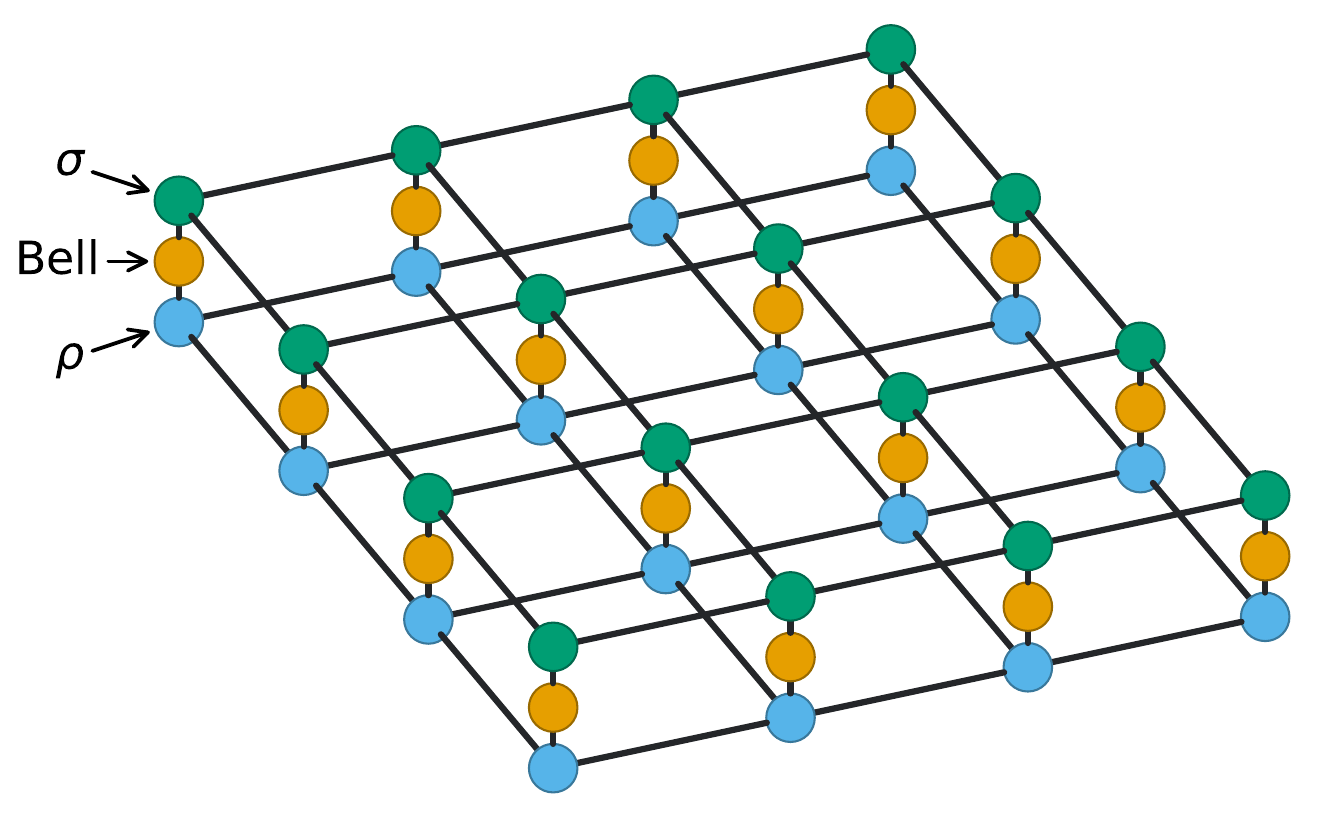}

\caption{Schematic illustration of fermionic $k$-RDM measurement by joint Bell measurements.\label{fig:joint-measurement}}
\end{figure}

Another way to implement fermionic shadows (but which is not limited to fermionic shadows) on quantum computers is to perform a joint measurement on two replicas of a given quantum state.
\cite{PhysRevLett.133.020601,PRXQuantum.6.010336} With some auxiliary (ancilla) state $\sigma$ that has the same size as $\rho$, one can perform a joint measurement of $O\otimes O$ on $\rho\otimes\sigma$ for all $4^{N}$ Pauli operators $O$, since all $O\otimes O$ are mutually commuting.
The joint measurement is performed again by the Bell pair measurement protocol, pairing the $i$th site of $\rho$ with the $i$th site of $\sigma$ for $i=1,\ldots,N$ (note that the pairing is different from the rainbow state pairing introduced in the previous section).
The obtained expectation values of $O\otimes O$ on the state $\rho\otimes\sigma$ relate to those of $O$ on $\rho$ via
\begin{equation}
\text{Tr}[O\rho]=\frac{\text{Tr}[(O\otimes O)(\rho\otimes\sigma)]}{\text{Tr}[O\sigma]}. \label{eq:bell_pauli}
\end{equation}
Unlike ordinary Bell sampling on $\rho\otimes\rho$, which estimates only squared expectation values $|\text{Tr}[O\rho]|^{2}$, the use of a known auxiliary $\sigma$ recovers signed linear estimators of $\text{Tr}[O\rho]$.
This construction specializes the $\rho\otimes\sigma$ identity used in the ``triply-efficient shadow tomography" framework of King et al.~\cite{PRXQuantum.6.010336} to tensor-network Hamiltonian estimation.
Note that this strategy is the same whether the systems (and operators) are bosonic or fermionic, as one can always apply the Jordan-Wigner transformation for fermions, and the generated non-local Pauli strings do not become a problem here.
Classically, the measurement can again be performed by sampling the product of Bell pair states $|\text{Bell}\rangle=\otimes_{i=1}^{N}|\text{bell}\rangle_{i_{\rho}i_{\sigma}}$, with the local Bell pairs $|\text{bell}\rangle_{i_{\rho}i_{\sigma}}$ updated sequentially using MCMC.
Here we choose $\sigma$ as a tensor network state, such that (1) $\text{Tr}[O\sigma]$ can be efficiently evaluated (by either an exact computation or the MC method), and (2) the overlap $\langle\text{Bell}|\rho\otimes\sigma|\text{Bell}\rangle$ can be computed with a double-layer contraction (see Fig.~\ref{fig:joint-measurement}).
If the bond dimension of $\sigma$ is $O(1)$ small, the evaluation of $\text{Tr}[O\sigma]$ can be viewed to have negligible cost, and double layer contraction effectively has the same cost as single-layer contraction (as one layer has very small bond dimension).

Although such a strategy seems to allow for the simultaneous measurement of all observables and give a significant improvement over other approaches, the real bottleneck comes from the small values of $\text{Tr}[O\sigma]$ that can appear in the denominator.
Specifically, the elements of fermionic $k$-RDMs cannot all be simultaneously large.
In fact, it has been shown that the Hilbert–Schmidt norm of the $k$-RDM is upper bounded by $O(N^{k/2})$.
\cite{visconti2026hilbert} In the Appendix~\ref{app:bell_rdm}, we show that, if the goal is to evaluate all $k$-RDM elements to a given additive error $\epsilon$, such a strategy has an $\Omega(N^{k}/\epsilon^{2})$ sample-complexity lower bound for arbitrary $\sigma$, up to logarithmic factors for simultaneous high-probability estimation. Therefore no asymptotic improvement is achieved (and in practice the method could even be worse since the state $\sigma$ induces extra cost and can be difficult to prepare).
We also note that the original proposed construction in~\cite{PRXQuantum.6.010336} using a mixed $\rho=\sigma$ and $\rho\neq \sigma$ scheme can also be similarly applied to classical TN simulations, giving a better $O(\log N)$ dependence, but with $O(1/\epsilon^4)$ error dependence, double-layer contraction to measure $\rho \otimes \rho$, and gives a biased estimator, therefore it is almost always much worse practically.

However, if we consider the expectation value of a linear function of the $k$-RDMs, this approach can be made powerful by choosing a good $\sigma$ state to utilize the given structure of the linear coefficients.
Consider the ab initio Hamiltonian in Eq.~\ref{eq:H_fermion}, and rewrite it in the Majorana basis as
\begin{equation}
H=\sum_{pq}\tilde{h}_{pq}\gamma_{p}\gamma_{q}+\sum_{pqrs}\tilde{V}_{pqrs}\gamma_{p}\gamma_{q}\gamma_{r}\gamma_{s}.\label{eq:H_majorana}
\end{equation}
Since Majorana operators map to Pauli operators after Jordan-Wigner transformation (up to a phase), one can evaluate

\begin{widetext}
\begin{align}
\text{Tr}[H\rho] & =\left\langle \sum_{pq}\frac{\tilde{h}_{pq}}{D_{pq}^{\sigma}}f(|\text{Bell}\rangle,\gamma_{p}\gamma_{q})+\sum_{pqrs}\frac{\tilde{V}_{pqrs}}{D_{pqrs}^{\sigma}}f(|\text{Bell}\rangle,\gamma_{p}\gamma_{q}\gamma_{r}\gamma_{s})\right\rangle _{\text{Bell}},\label{eq:bell_estimator}
\end{align}
\end{widetext}
where $f(|\text{Bell}\rangle,O)=\langle\text{Bell}|O\otimes O|\text{Bell}\rangle$, $D_{pq}^{\sigma}=\text{Tr}[\gamma_{p}\gamma_{q}\sigma]$ and $D_{pqrs}^{\sigma}=\text{Tr}[\gamma_{p}\gamma_{q}\gamma_{r}\gamma_{s}\sigma]$ are the Majorana $1$-RDM and $2$-RDM, respectively.
In ab initio quantum chemistry Hamiltonians, the Coulomb interaction expressed in exponentially localized bases (such as Gaussian or atomic bases) has exponentially decaying dependence on $r_{pr}$ and $r_{qs}$: $V_{pqrs}\sim\exp(-cr_{pr})\exp(-cr_{qs})$.
Motivated by this, we choose the $\sigma$ state as a Slater determinant, whose $2$-RDM is related to the $1$-RDM by Wick's theorem (note, this is in the Majorana basis) 
\begin{equation}
D_{pqrs}=D_{pq}D_{rs}-D_{pr}D_{qs}+D_{ps}D_{qr} \label{eq:wick}
\end{equation}
We then choose the $1$-RDM elements to have exponentially decaying structure $D_{pq}^{\sigma}\sim A\exp(-c^{\prime}r_{pq})$, but with a smaller exponent $c^{\prime}<c$.
Such a structure automatically ensures $A\sim O(1)$, which implies that $\sigma$ can be written as a tensor network with low bond dimension.
Finally  we expect that for $p, q, r, s$ all distinct, $\frac{\tilde{V}_{pqrs}}{D_{pqrs}^{\sigma}}$ has the same exponential distance-decaying structure with $\tilde{V}_{pqrs}$, only with a lower exponential exponent. Combined with physical assumptions about the quantum state, such as rapidly decaying correlations and a neutral system with an extensive Coulomb energy, we then expect the variance to be reduced to $O(N)$.

\subsubsection{Environment reuse strategy} \label{subsec:reuse}
A common trick in tensor network algorithms is to reuse the environment for multiple contractions. This reuse is exact in 1D tensor network algorithms, and is approximate for higher-dimensional contractions.
For any given scalar-output tensor network contraction, one can effectively construct all the local environments (e.g. around each site) during the contraction process, thus such local environments can be reused for other scalar-output tensor network contractions that are only locally different.
Such reuse can be applied to direct expectation value evaluation in direct contraction, MCMC sampling, and local observable evaluation in the standard MC approach, so it gives $O(N)$ cost reduction in all the approaches (direct contraction, standard MC or shadow estimators) discussed above.
Therefore, it does not affect the relative improvements we obtain.
Details of how the $O(N)$ reduction is achieved in these approaches, is discussed in Appendix~\ref{app:reuse}.

\subsubsection{Variational ground state simulation} \label{subsec:variational}
One of the most important applications of evaluating expectation values of Hamiltonians is in variational ground state simulation. This is the main target of variational Monte Carlo. However, an additional consideration when comparing the shadow tomography approaches with standard variational Monte Carlo is the zero-variance principle (ZVP) \cite{PhysRevLett.83.4682} in standard MC evaluation. Consider the form of the local observable in Eq.~\ref{eq:standard_MC}, let $O$ be the Hamiltonian, then $O_{\text{loc}}^{\text{standard}}(x)$ becomes the corresponding energy eigenvalue which is independent of $x$, if $|\psi\rangle$ is already the ground state (or any eigenstate), thus the sampling variance becomes zero. However, the shadow estimators introduced above do not have this property.

Nevertheless, the loss of the ZVP can be partly compensated by an additional efficient approximation when evaluating energy gradients w.r.t. parameters (tensor values). Consider the tensor network state $|\psi(\boldsymbol{\theta})\rangle$ parameterized by tensor values $\boldsymbol{\theta}=\{\boldsymbol{\theta}_{1},...,\boldsymbol{\theta}_{N}\}$, where $\boldsymbol{\theta}_{i}$ is the set of tensor values on the $i$th site, and decompose the energy expectation value into contributions from local sites, which are evaluated by MCMC sampling (either using standard MC or shadow estimators) $E=\sum_{j}E_{j}=\left\langle E_{j}(n)\right\rangle _{n}$. The energy gradient can be written as 
\begin{equation}
\partial_{\boldsymbol{\theta}_{i}}E=\sum_{j}\partial_{\boldsymbol{\theta}_{i}}E_{j}\label{eq:gradient_decomposition}
\end{equation}
In states with fast decaying correlations, we have $\partial_{\boldsymbol{\theta}_{i}}E_{j}\sim\exp(-r_{ij})$, where $r_{ij}$ is the distance between site $i$ and $j$. However, the Monte Carlo evaluation of $\partial_{\boldsymbol{\theta}_{i}}E_{j}$ contributes an $O(1)$ variance regardless of the separation of $i$ and $j$, which contributes to an unphysical $O(N)$ higher sampling variance for the energy gradient. Such an unphysical variance contribution can be removed by neglecting distant $(i,j)$ pairs, but such an approximation breaks the ZVP, and is thus not compatible with the standard VMC approach (unless the ZVP is abandoned). Note that such a strategy is valid for all the three shadow estimators introduced above and can be used even if the correlations decay algebraically. To summarize, for the task of variational ground state simulation, the standard VMC algorithm benefits from the ZVP, but for many relevant physical states, shadow estimators can benefit from removing the unphysical sampling variance contribution by the above truncation. Further details are discussed in Appendix~\ref{app:grad_trunc}.

\begin{figure*}[t]
\centering
\includegraphics[width=0.98\linewidth]{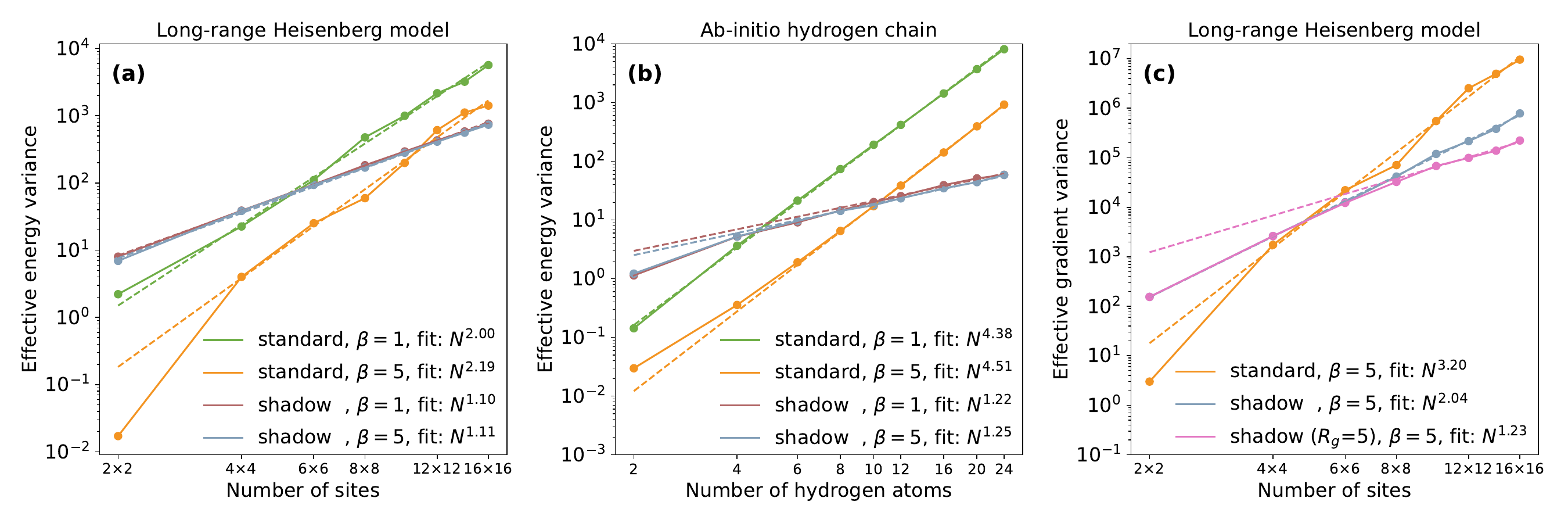}

\caption{Scaling of the effective energy and energy-gradient variance of TN states for the long-range Heisenberg model (bosonic) and ab-initio hydrogen chain (fermionic) using the standard MC and shadow estimators.
(a) Energy variance of the long-range Heisenberg model. (b) Energy variance of the ab-initio hydrogen chain. (c) Energy-gradient variance of the long-range Heisenberg model, including the Pauli-shadow estimator with gradient truncation distance $R_{g}=5$.
The considered shadow estimators are the Pauli shadow for the Heisenberg model, and the Bell-sampling shadow for the ab-initio hydrogen chain.
Tensor network states use bond dimension $D=4$ for the 2D Heisenberg model, and $D=16$ for the ab-initio hydrogen chain.
For both systems, the variances are estimated for states obtained by approximate imaginary-time evolution from an initial product state, $e^{-\beta H}|\phi_{0}\rangle$. Panels (a) and (b) compare $\beta=1$ and $\beta=5$, representing states farther from and closer to Hamiltonian eigenstates, while panel (c) shows the gradient variance at $\beta=5$.
The numerically fitted scalings are in good agreement with the theoretical scalings summarized in Table~\ref{table:scaling}, after properly accounting for the environment reuse and converting between absolute and relative errors.
\label{fig:var_scaling}}
\end{figure*}

\subsubsection{Stability of the gradient estimator} \label{subsec:stability}
We next discuss a different stability property of the gradient estimator in the variational optimization: the shadow gradient estimator has a bounded variance if the gradient of the parameterized state itself $\partial_{\theta}|\psi(\theta)\rangle$ has a bounded norm, while the standard MC estimator does not have such an upper bound, and thus can suffer from stability issues. 

For both the standard MC estimator and the shadow estimator, the energy gradient estimator is

\begin{equation}
\begin{aligned}
\partial_{\theta}E(\theta)
&=\mathbb{E}_{x\sim p_{\theta}(\cdot)}
\left[g_{\text{loc}}(x)\right],\\
g_{\text{loc}}(x)
&=\left(E_{\mathrm{loc}}(x)-E\right)
\partial_{\theta}\log p_{\theta}(x),
\end{aligned}
\label{eq:standard_gradient}
\end{equation}
and the difference is just (1) the sampling ensemble of $x$, and (2) the form of $E_{\text{loc}}(x)$. We prove in Appendix \ref{app:stability_proof} that, if the sampling ensemble of $x$ is a local 1-design (i.e. satisfying the local resolution of the identity), we have
\begin{equation}
\begin{aligned}
\text{Var}\left[g_{\text{loc}}(x)\right]
&\leq
16\Vert\partial_{\theta}|\psi(\theta)\rangle\Vert_{2}^{2}
\max_{x}|E_{\mathrm{loc}}(x)|^{2}.
\end{aligned}
\end{equation}

For the standard MC estimator, $E_{\text{loc}}(x)$ is given in Eq. \ref{eq:standard_MC}, which is generally unbounded. In the Appendix \ref{app:stability_counterexample}, we present a concrete example where $\text{Var}\left[g_{\text{loc}}(x)\right]$ can be unbounded, even if $\Vert\partial_{\theta}|\psi(\theta)\rangle\Vert_{2}$ is bounded.
The numerical stability of gradient estimators is a well-known issue in the general VMC literature, and generally results in more difficult optimizations~\cite{pathak2020light,chen2022proximal}.

For the three types of shadow estimators mentioned in this work, we show that they all give bounded gradient variance. For Pauli shadows, the local observable is
\begin{equation}
O_{\mathrm{loc}}^{\text{Pauli}}(x)=\text{Tr}[\rho_{s}(U,x)H]
\end{equation}
Decomposing $H$ to Pauli operators $O$, we have $|O_{\mathrm{loc}}^{\text{Pauli}}(x)|\leq3^{k}$, therefore $E_{\mathrm{loc}}(x)$ is also bounded. The rainbow-basis shadow and Bell sampling shadow both fall under the commuting measurements framework. According to Eq. \ref{eq:commuting_evaluation}, in the mutually diagonalized basis $x$ of the commuting observables, the Pauli (or Majorana) observables simply give
\begin{equation}
O_{\mathrm{loc}}^{\text{commuting}}(x)=\langle x|O|x\rangle=\pm1,
\end{equation}
which therefore is again bounded. We also note that in the Bell sampling shadow, the true expectation value requires dividing by $\text{Tr}[O\sigma]$ in Eq. \ref{eq:bell_pauli}, which is another possible source of unbounded gradient variance behavior. However, as discussed in Sec. \ref{subsec:bell}, this may be avoided by a careful choice of the state $\sigma$.

\section{Numerical results}

\subsection{Energy evaluation in a long-range Heisenberg model and for the ab initio hydrogen
chain}

To quantitatively test the variance scaling predicted in the theory section, we consider two representative systems covering both spin and fermionic cases.
The first is the transverse-field long-range Heisenberg model on a 2D square lattice with open boundary conditions, 
\begin{equation}
H=-J\sum_{ij}r_{ij}^{-\alpha}\boldsymbol{S}_{i}\cdot\boldsymbol{S}_{j}-h\sum_{i}S_{i}^{x}, 
\end{equation}
with parameters $\alpha=3$ and $J=h=-1$.
The transverse field breaks $S_{z}$ symmetry, so that a simple Monte Carlo sampling move of flipping individual spins can explore the full Hilbert space without ergodicity issues.
The second system is an ab initio one-dimensional hydrogen chain (i.e. fermionic) \cite{PhysRevX.7.031059} in a minimal basis with an equally spaced bond length $1.4$ \AA, where the Hamiltonian is given in Eq.~\ref{eq:H_fermion} with ab initio derived integrals.

For both systems, we construct tensor network states by approximate imaginary-time evolution from some initial product state $|\phi_{0}\rangle$, namely $e^{-\beta H}|\phi_{0}\rangle$, with two representative imaginary times $\beta=1$ and $\beta=5$.
These two choices provide states respectively farther from and closer to Hamiltonian eigenstates, allowing us to directly probe how the variance depends on the closeness of the input state to an eigenstate.

Here we focus on comparisons between shadow estimators and the standard MC approaches, where the total costs are all given by Eq.
\ref{eq:obs_var}.
Since the single-contraction cost and the $1/\epsilon^{2}$ dependence are the same, we focus on the remaining part, and accordingly define the ``effective variance'' as
\begin{equation}
\begin{aligned}
\text{effective variance} & =(\text{variance of evaluation})\\
 & \times(\text{TN contractions per evaluation}).
\end{aligned}
\end{equation}
For gradient estimation, we define the gradient variance as the sum of the variances of all tensor-parameter components of the gradient estimator, and use the same effective-variance definition after multiplying by the TN contractions per gradient evaluation.
The environment reuse strategy discussed in Sec.~\ref{subsec:reuse} is applied to all computations and reflected in the reported results (giving $O(N)$ improvements over all reported scalings).

Figure~\ref{fig:var_scaling} shows the effective variance of both energy and gradient estimators as a function of system size.
Panels (a) and (b) compare energy estimation for the long-range Heisenberg model and the ab-initio hydrogen chain, respectively, while panel (c) compares gradient estimation for the long-range Heisenberg model.

For the long-range Heisenberg model {[}Fig.~\ref{fig:var_scaling}(a){]}, the standard Monte Carlo estimator exhibits an effective variance scaling close to $O(N^{2})$, for both $\beta=1$ and $\beta=5$.
By contrast, the Pauli shadow estimator scales approximately as $O(N)$ for both $\beta=1$ and $\beta=5$, in good agreement with the theoretical prediction.
This improvement comes from the $O(N)$ reduction in terms of TN contractions per evaluation, while the scalings with respect to the variance of evaluation are the same.
In addition, the two Pauli shadow curves are nearly identical, whereas the standard MC sampling results show a significantly reduced variance for $\beta=5$ compared to $\beta=1$, which reflects the effect of the ZVP.

For the ab-initio hydrogen chain {[}Fig.~\ref{fig:var_scaling}(b){]}, the same pattern becomes even more pronounced.
The standard Monte Carlo estimator scales approximately as $O(N^{4})$ for both $\beta=1$ and $\beta=5$, due to the $O(N^{4})$ number of terms in the Hamiltonian.
In contrast, the effective variance of the Bell sampling strategy scales as only $\sim O(N)$ for $\beta=1$ and $\beta=5$, again with almost identical behavior for the two states.
In the fermionic case as well, the standard MC method retains its familiar advantage that the variance becomes smaller when the input state is closer to an eigenstate $(\beta=5)$, while the shadow estimator shows much weaker dependence on this state property but a dramatically improved scaling with system size.
As a result, although the standard Monte Carlo estimator can be competitive for smaller systems or states close to eigenstates, the superior asymptotic scaling of the shadow estimator makes it more favorable in the large-system regime.
Finally, we also note that, although we only compare Monte Carlo methods, the same scaling improvement is also expected when comparing against the standard ab initio DMRG direct contraction algorithm.

\begin{figure}
\centering
\includegraphics[width=0.75\columnwidth]{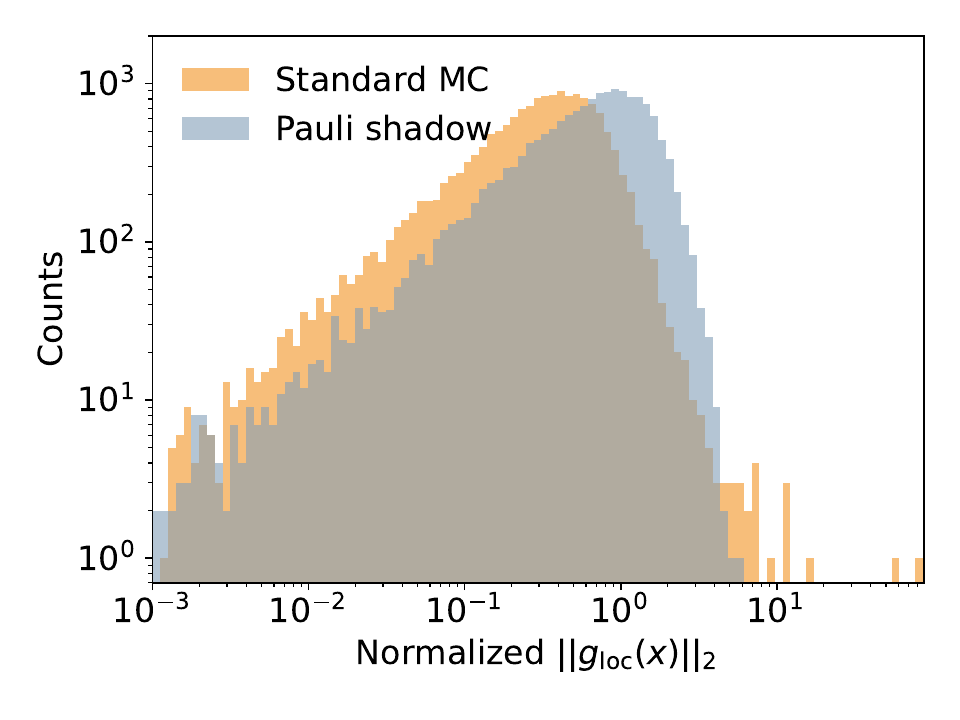}

\caption{Histogram of the normalized local gradient-estimator norm $\|g_{\text{loc}}(x)\|_2$ for the standard MC and Pauli shadow estimators in the $16\times16$ long-range Heisenberg model.
The samples $x$ are drawn from the corresponding Born distributions $p(x)=|\langle x | \psi \rangle |^2$, using a state taken from one step of the variational ground-state optimization.
For each estimator, $\|g_{\text{loc}}(x)\|_2$ is normalized by its standard deviation to compare the relative shapes of the distributions.}
\label{fig:grad_dist}
\end{figure}

Figure~\ref{fig:var_scaling}(c) further shows the effective variance of the energy-gradient estimator for the long-range Heisenberg model at $\beta=5$.
Without gradient truncation, the standard MC gradient estimator scales approximately as $O(N^{3})$, while the Pauli-shadow gradient estimator scales closer to $O(N^{2})$.
This shows that the scaling advantage of the shadow estimator is not limited to scalar energy evaluation, but also carries over to the stochastic gradients used in variational optimization.
The gradient-variance scaling without truncation is one power of $N$ higher than the corresponding energy-variance scaling for both standard MC and the Pauli shadow.
As discussed in Sec.~\ref{subsec:variational}, this extra factor is an unphysical contribution from summing noisy distant local-gradient terms.
After applying a truncation of gradient terms (see Eq.~\ref{eq:gradient_decomposition}) with $R_{\mathrm{g}}=5$, the Pauli-shadow gradient variance is reduced to a scaling close to $O(N)$.
The fitted exponent $1.23$ is slightly larger than its asymptotic expected value, which we attribute to the limited system size not yet fully reaching the large-$N$ regime.
The crossing point where the Pauli-shadow gradient estimator becomes more efficient than standard MC also occurs at a smaller system size than the corresponding crossing in the energy variance, which is a favorable sign for variational optimization because the gradient, rather than the energy estimator alone, directly controls the parameter updates.

Overall, the shadow estimator achieves approximately $O(N)$ effective variance scaling for both the long-range Heisenberg model (numerically around $O(N^{1.1})$) and ab-initio hydrogen chain (numerically around $O(N^{1.2})$).
Multiplied by the $O(N)$ TN contraction complexity, it gives $O(N^2)$ total complexity to reach an arbitrarily fixed absolute (additive) error, or equivalently $O(1)$ total complexity for an arbitrarily fixed relative (multiplicative) error, assuming the total energy is extensive.
Such optimal $O(1)$ scalings, as well as the scalings of the standard Monte Carlo estimators, match the theoretical scalings summarized in Table~\ref{table:scaling} (after including the environment reuse, and converting properly between absolute and relative errors).

\begin{figure*}
\centering
\includegraphics[width=1\linewidth]{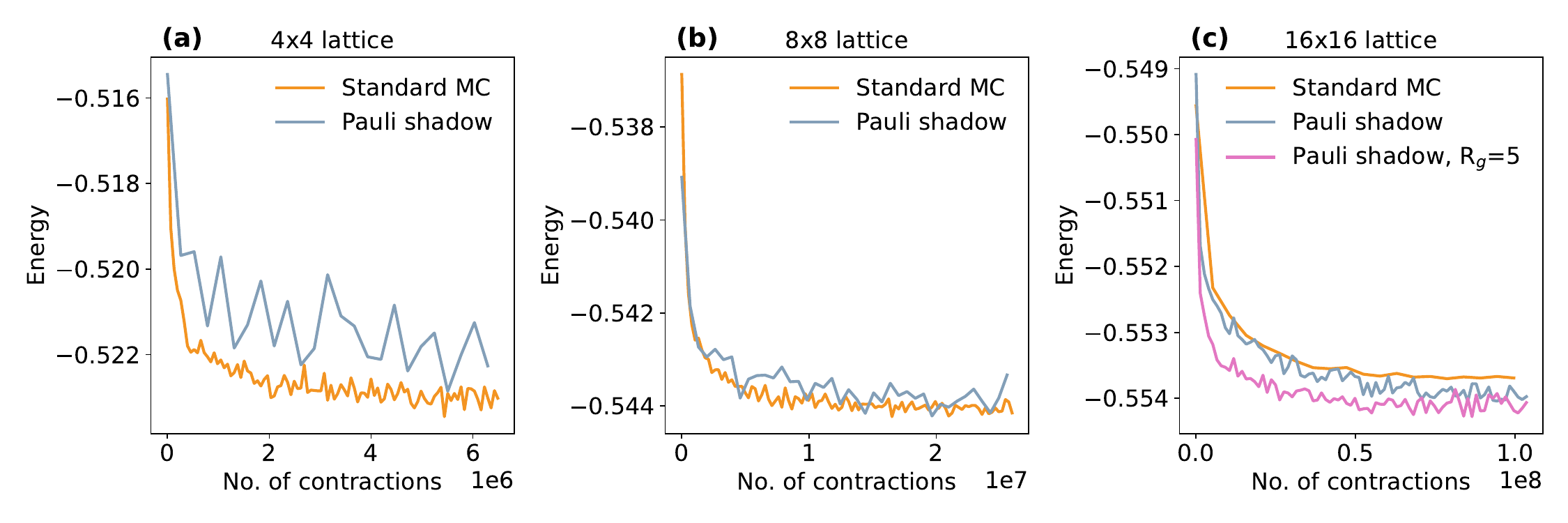}

\caption{Variational tensor network simulation of the ground state of a 2D long-range antiferromagnetic Heisenberg model on the square lattice, by the standard Monte Carlo strategy (blue), and the Pauli shadow strategy (orange) proposed in this work.
All simulations use bond dimension $D=4$, the same gradient-based optimizer and learning rate schedule.
The same number of single-layer contractions are used for both the standard MC and Pauli shadow.
The Pauli shadow combined with the gradient truncation scheme with distance $R_{\text{g}}=5$ is also shown (green) for the simulation of the $16\times16$ lattice.\label{fig:heisenberg}}
\end{figure*}

\subsection{Variational ground state simulation}

While the previous subsection focuses on the scaling of energy evaluation for fixed tensor network states, the practically more important task is variational ground-state simulation, where these estimators are used repeatedly throughout an optimization process.
In this setting, we expect that standard MC benefits from the zero-variance principle near eigenstates, whereas the Pauli shadow approach has the advantage of improved scaling with system size, and can be further combined with the gradient truncation strategy discussed above for even better scaling.

We compare the Pauli shadow based MC strategy and the standard MC strategy on three different lattice sizes $4\times4$, $8\times8$ and $16\times16$, with a fixed 2D tensor network bond dimension $D=4$.
We use the Adam optimizer, fix the learning rate schedule as exponentially decaying from 0.03 to 0.0003, and use the same total number of single-layer contractions for the two strategies.

The improved gradient stability can be seen directly from the distribution of the local gradient estimator.
As shown in Fig.~\ref{fig:grad_dist}, after normalizing by the standard deviation, the standard MC estimator exhibits a broader distribution of $\|g_{\mathrm{loc}}(x)\|_2$, including a heavier tail of rare large-gradient samples.
The Pauli-shadow estimator gives a more concentrated distribution, consistent with the bounded-variance stability analysis in Sec.~\ref{subsec:stability}.
This provides a microscopic explanation for the smoother large-system optimization behavior observed below.

The variational energy as a function of the number of evaluated single-layer contractions for the three lattices is shown in Fig.
\ref{fig:heisenberg}.
We observe a clear transition in the relative performance of the two strategies as the system size increases from $4\times4$ to $16\times16$.
For the small $4\times4$ lattice, the standard MC approach exhibits faster energy convergence and lower fluctuations.
This is expected behavior as the standard approach benefits from the zero-variance principle (ZVP).
However, as the system size increases, the scaling advantage of the Pauli shadow MC gradually appears.
For the $8\times8$ lattice, Pauli shadow MC shows similar energy convergence as the standard MC; while for the $16\times16$ lattice, the Pauli shadow already shows significantly better energy convergence.
Further improvement is observed when the Pauli shadow strategy is combined with the gradient truncation scheme, with the truncation distance $R_{\text{g}}=5$ (i.e. $\partial_{\boldsymbol{\theta}_{i}}E_{j}$ is neglected for $(i,j)$ pairs with distance larger than $R_{\text{g}}$), which evaluates the gradients with much lower variance and negligible truncation error.
Here gradient truncation is applied only to the optimization gradients, while the energy estimator used to report the variational energy remains the full unbiased estimator.
Therefore, the lower optimized energy obtained with gradient truncation indicates a genuine improvement of the optimized state; at minimum, it shows that the benefit of reduced gradient noise outweighs the truncation error in the gradient direction.
This suggests that for large-scale 2D simulations where the number of terms in the Hamiltonian is large, the Pauli shadow strategy also outperforms existing MC algorithms in the systems of practical interest.

\section{Conclusion and outlook}
In this work, we have adapted ideas from shadow tomography, which has been mainly developed in the quantum information community, to the problem of observable evaluation in classical tensor network simulations. The central observation is that methods such as Pauli shadows and Bell-basis measurements allow a single measurement outcome to provide information about many observables simultaneously. While these ideas were originally designed for quantum experiments or quantum computers, we showed that, with suitable modifications, the same principle can be implemented in classical tensor network Monte Carlo algorithms using only tensor-network amplitude evaluations.

For spin systems, directly applying the standard Pauli-shadow construction gives an $O(N)$ improvement for complete fixed-$k$ RDM estimation, and reduces the cost of long-range two-body Hamiltonian expectation value estimation by a factor of $O(N)$ compared with standard Monte Carlo. For fermionic systems, ordinary Pauli shadows are not directly compatible with tensor network locality because of nonlocal Jordan--Wigner strings. We therefore introduced a rainbow-basis shadow construction based on mutually commuting Majorana groupings, giving an $O(N)$ improvement for complete fermionic 2-RDM estimation up to logarithmic factors. For scalar ab initio Hamiltonian expectation values, we adapted the Bell-sampling strategy with an auxiliary state $\sigma$, reducing the scaling by a factor of $O(N^{3})$ compared with standard Monte Carlo under physically motivated locality assumptions. With the standard environment reuse strategy, shadow estimators achieve the optimal $O(1)$ overall sample scaling for evaluating the expectation values of both spin and fermionic Hamiltonians to an arbitrarily fixed Monte Carlo relative error.

We further showed that these shadow estimators can be useful in variational tensor network optimization. Although they do not satisfy the zero-variance principle, they have better large-system scaling and can provide more stable gradient estimators. Combined with gradient truncation, this leads to improved performance and accuracy in the large-system regime, as demonstrated in long-range Heisenberg models and ab initio hydrogen chains. An important future direction is to optimize the choice of the auxiliary state $\sigma$ in the Bell-sampling construction, especially for more complicated and higher-dimensional ab initio systems. More broadly, our results suggest that quantum-information tools can provide useful new classical algorithms for tensor network simulations.

\section{Acknowledgement}
This work is supported by the US Department of Energy, Office of Science, Accelerated Research in Quantum Computing Centers, Quantum Utility through Advanced Computational Quantum Algorithms,
through Award No. DE-SC0025572. We also thank Prof. Hsin-Yuan Huang, Dr. Johnnie Gray, Jielun Chen, Gunhee Park and Sijing Du for helpful discussions and comments.
\bibliography{main}

\section{Appendix}

\subsection{2RDM estimation using the rainbow-basis shadow}\label{app:rainbow_2rdm}
The grouping argument in this Appendix is stated for a fixed Majorana flavor pattern.
We write the two Majoranas associated with site $i$ as $\gamma_{i,p}\equiv\gamma_{2i-p}$, with $p\in\{0,1\}$.
A fermionic $2$-RDM element expands into at most $2^{4}$ Majorana strings with fixed choices of the four binary flavor labels.
We apply the grouping construction below to each fixed flavor pattern separately.
To define the actual measurement basis, each measured Majorana pair $\gamma_{i,p}\gamma_{j,q}$ is completed by the complementary pair $\gamma_{i,1-p}\gamma_{j,1-q}$ on the same two physical sites.
This gives a product of two-site Bell-pair bases, together with local occupation bases on unpaired sites.
The fixed-flavor construction therefore gives only a constant-factor overhead and does not affect the asymptotic scaling.

Consider a fixed-flavor Majorana product $\gamma_{a,p_a}\gamma_{b,p_b}\gamma_{c,p_c}\gamma_{d,p_d}$ with site indices $a<b<c<d$. Let $R_{ij}=\frac{1}{2}(R_{i}+R_{j})$, $r_{ij}=\frac{1}{2}(R_{j}-R_{i})$. Without loss of generality, we assume $r_{ab}\leq r_{cd}$; otherwise one just needs to reverse the order of $(a,b,c,d)$. Two grouping strategies mentioned in the main text are:

(1): a group with fixed rainbow centers $R_{ab}$ and $R_{cd}$, with rainbow widths $r_{ab}$ and $r_{cd}$ satisfying 
\begin{equation}
\{r_{ab}\leq2^{m_{1}},r_{cd}<(R_{cd}-R_{ab})-2^{m_{1}}\}
\end{equation}
 for a given integer $m_{1}$.

(2): a group with fixed rainbow centers $R_{bc}$ and $R_{ad}$, with rainbow widths $r_{bc}$ and $r_{ad}$ satisfying 
\begin{equation}
\{r_{bc}\leq2^{m_{2}},r_{ad}>(R_{ad}-R_{bc})+2^{m_{2}}\}
\end{equation}
 for a given integer $m_{2}$.

We first verify that each group contains mutually commuting Majorana products. To do this, we only need to verify that the two rainbows have no overlap. For class (1), the non-overlapping condition is 
\begin{equation}
R_{ab}+r_{ab}<R_{cd}-r_{cd}
\end{equation}
which is automatically satisfied. For class (2), the assumption $r_{cd}\geq r_{ab}$ is equivalent to $R_{ad}\geq R_{bc}$. Therefore, the non-overlapping condition is
\begin{equation}
R_{ad}-r_{ad}<R_{bc}-r_{bc}
\end{equation}
which is again automatically satisfied. The other containment condition, $R_{bc}+r_{bc}<R_{ad}+r_{ad}$, is also automatically satisfied.

Next we verify that all fixed-flavor Majorana products must fall into one of the groups. Choose $m_{1}$ such that
\begin{equation}
2^{m_{1}-1}<r_{ab}\leq2^{m_{1}}.
\end{equation}
If the operator is not contained in the corresponding group in the first class, then it must satisfy
\begin{equation}
r_{cd}\geq(R_{cd}-R_{ab})-2^{m_{1}}.
\end{equation}
This condition is equivalent to 

\begin{equation}
\begin{aligned} & r_{cd}\geq(r_{ab}+2r_{bc}+r_{cd})-2^{m_{1}}\\
\Leftrightarrow & r_{ab}+2r_{bc}\leq2^{m_{1}}.
\end{aligned}
\end{equation}
Therefore we must have
\begin{equation}
2^{m_{1}-1}<r_{ab}\leq r_{ab}+2r_{bc}\leq2^{m_{1}},
\end{equation}
which gives 
\begin{equation}
r_{bc}<2^{m_{1}-2}<\frac{1}{2}r_{ab}.\label{eq:exception1}
\end{equation}

Similarly, choose $m_{2}$ such that
\begin{equation}
2^{m_{2}-1}<r_{bc}\leq2^{m_{2}}.
\end{equation}
If the operator is not contained in the corresponding group in the second class, then it must satisfy
\begin{equation}
r_{ad}\leq(R_{ad}-R_{bc})+2^{m_{2}}.
\end{equation}
This condition is equivalent to

\begin{equation}
\begin{aligned} & r_{ab}+r_{bc}+r_{cd}\leq R_{ad}-R_{bc}+2^{m_{2}}=r_{cd}-r_{ab}+2^{m_{2}}\\
\Leftrightarrow & r_{bc}+2r_{ab}\leq2^{m_{2}}.
\end{aligned}
\end{equation}
Therefore we must have
\begin{equation}
2^{m_{2}-1}<r_{bc}\leq r_{bc}+2r_{ab}\leq2^{m_{2}},
\end{equation}
which gives 
\begin{equation}
r_{ab}<2^{m_{2}-2}<\frac{1}{2}r_{bc}.\label{eq:exception2}
\end{equation}

Clearly the two conditions in Eq.~\ref{eq:exception1} and Eq.~\ref{eq:exception2} cannot both be true. Therefore, every fixed-flavor quartic Majorana product is covered by at least one of the two grouping strategies.
After summing over the $2^{4}$ fixed flavor patterns, all fermionic $2$-RDM elements are covered with only a constant-factor overhead.
With this simplified presentation, different flavor patterns and their corresponding Bell-pair completions are not combined into the same group, so each group contains only a constant fraction of the operators that could be included in an optimized implementation.
This constant-factor inefficiency can be removed by merging compatible flavor patterns and choosing the Bell-pair completions jointly, but we do not present this optimization since it does not affect the scaling.

\subsection{RDM estimation using the Bell sampling shadow}\label{app:bell_rdm}
We show that Bell sampling does not improve the asymptotic sample complexity for estimating all elements of a complete fermionic $k$-RDM to additive error.
Let $\{O_{\mu}\}_{\mu=1}^{M_{k}}$ denote the $2k$-Majorana products appearing in the complete $k$-RDM, where $M_{k}=\Theta(N^{2k})$.
For Bell sampling with auxiliary state $\sigma$, the estimator for $O_{\mu}$ has the form
\begin{equation}
\widehat{O}_{\mu}=\frac{f_{\mu}}{D_{\mu}^{\sigma}},
\qquad
D_{\mu}^{\sigma}=\operatorname{Tr}[O_{\mu}\sigma].
\end{equation}
Here $f_{\mu}=f(|\mathrm{Bell}\rangle,O_{\mu})$ is the numerator obtained from the Bell-basis measurement.
For Majorana products, the numerator has unit second moment,
\begin{equation}
\mathbb{E}[|f_{\mu}|^{2}]=1.
\end{equation}
Let
\begin{equation}
m_{\mu}=\operatorname{Tr}[O_{\mu}\rho]
\end{equation}
be the true expectation value.
Unbiasedness of the Bell-sampling estimator gives
\begin{equation}
\mathbb{E}[f_{\mu}]=D_{\mu}^{\sigma}m_{\mu}.
\end{equation}
Therefore
\begin{equation}
\operatorname{Var}[\widehat{O}_{\mu}]
=
\frac{\operatorname{Var}[f_{\mu}]}{|D_{\mu}^{\sigma}|^{2}}
=
\frac{1-|D_{\mu}^{\sigma}m_{\mu}|^{2}}{|D_{\mu}^{\sigma}|^{2}}.
\end{equation}

The proof is based on comparing the total numerator variance with the total denominator weight.
We use the $k$-RDM norm bound of Ref.~\cite{visconti2026hilbert}.
Although this bound is stated for the ordinary fermionic $k$-particle RDM, the same scaling applies to the Majorana $k$-RDM used here, since changing between creation-annihilation operators and Majorana operators is a fixed local linear transformation for fixed $k$.
Thus, for any state $\tau$,
\begin{equation}
\sum_{\mu=1}^{M_{k}}|\operatorname{Tr}[O_{\mu}\tau]|^{2}
\leq C_{k}N^{k},
\end{equation}
where $C_{k}$ is independent of $N$.
Applying this bound to $\tau=\rho$ and $\tau=\sigma$ gives
\begin{equation}
\sum_{\mu=1}^{M_{k}}|m_{\mu}|^{2}=O(N^{k}),
\qquad
\sum_{\mu=1}^{M_{k}}|D_{\mu}^{\sigma}|^{2}=O(N^{k}).
\end{equation}

The first bound shows that the true-value subtraction in the numerator variance is negligible in the total scaling.
Indeed, since $|D_{\mu}^{\sigma}|\leq1$,
\begin{equation}
\begin{aligned}
\sum_{\mu=1}^{M_{k}}\operatorname{Var}[f_{\mu}]
&=
\sum_{\mu=1}^{M_{k}}\left(1-|D_{\mu}^{\sigma}m_{\mu}|^{2}\right)\\
&\geq
M_{k}-\sum_{\mu=1}^{M_{k}}|m_{\mu}|^{2}\\
&=\Omega(N^{2k}).
\end{aligned}
\end{equation}
The second bound gives the total denominator weight,
\begin{equation}
\sum_{\mu=1}^{M_{k}}|D_{\mu}^{\sigma}|^{2}=O(N^{k}).
\end{equation}

Finally, using a weighted-average bound,
\begin{equation}
\begin{aligned}
\max_{\mu}\operatorname{Var}[\widehat{O}_{\mu}]
&=
\max_{\mu}\frac{\operatorname{Var}[f_{\mu}]}{|D_{\mu}^{\sigma}|^{2}}\\
&\geq
\frac{\sum_{\mu}\operatorname{Var}[f_{\mu}]}
{\sum_{\mu}|D_{\mu}^{\sigma}|^{2}}\\
&=\Omega(N^{k}).
\end{aligned}
\end{equation}
Therefore at least one $k$-RDM element has Bell-sampling variance $\Omega(N^{k})$.
Consequently, estimating all $k$-RDM elements to additive error $\epsilon$ requires
\begin{equation}
\Omega\left(\frac{N^{k}}{\epsilon^{2}}\right)
\end{equation}
samples, up to logarithmic factors for simultaneous high-probability estimation of all elements.

\subsection{Environment reuse}\label{app:reuse}
Environment reuse is a standard contraction-reuse idea in tensor-network algorithms.
For simplicity, we describe it first in the MPS case.
The main purpose of this Appendix is to clarify how the per-sample tensor-network cost is decomposed for the standard MC and shadow estimators.
For both methods, a sample is generated by Metropolis sampling with sequential local updates.
After a sample is obtained, one evaluates a local observable on that sample.
The two methods therefore share the same sampling cost, but differ in the local-observable cost: standard MC requires additional tensor-network amplitudes, while the shadow estimator computes the sampled observable directly from the measurement outcome.

For a fixed configuration \(x=(x_1,\ldots,x_N)\), the MPS amplitude is a one-dimensional chain contraction,
\begin{equation}
W(x)=\langle x|\psi\rangle=A^{x_1}_1 A^{x_2}_2\cdots A^{x_N}_N .
\end{equation}
After the physical indices are fixed by \(x\), there are only virtual bonds left to contract.
The standard environment-reuse construction is to build left and right environments,
\begin{equation}
\begin{aligned}
L_i(x) &= A^{x_1}_1\cdots A^{x_{i-1}}_{i-1},\\
R_i(x) &= A^{x_{i+1}}_{i+1}\cdots A^{x_N}_N.
\end{aligned}
\end{equation}
Then a one-site modification is evaluated as
\begin{equation}
W(x_1,\ldots,x_i',\ldots,x_N)=L_i(x)A_i^{x_i'}R_i(x),
\end{equation}
and a nearest-neighbor modification is evaluated similarly as
\begin{equation}
W(\ldots,x_i',x_{i+1}',\ldots)=L_i(x)A_i^{x_i'}A_{i+1}^{x_{i+1}'}R_{i+1}(x).
\end{equation}
Thus, once the environments are available, locally modified amplitudes are obtained by local contractions rather than by recomputing the full chain.

We first apply this to the local-observable evaluation in standard MC.
For a sampled configuration \(x\), the local energy requires many amplitudes \(\langle x'|\psi\rangle\), where \(x'=O x\) for different Hamiltonian terms \(O\).
When \(O\) changes only one site, or a few nearby sites, \(x'\) differs from \(x\) only locally.
The environments above allow all such locally modified amplitudes to be evaluated by local contractions.
In contrast, for a shadow estimator, once the sampled measurement outcome is obtained, the local shadow value is computed directly from that outcome.
No additional tensor-network amplitude contractions are needed for the local observable.

A closely related reuse applies to direct contraction of observables.
For \(\langle\psi|O|\psi\rangle\), one first contracts the physical bonds of the double-layer tensor network.
For a fixed local operator insertion, this produces a tensor network with only virtual bonds left to contract.
If many observables differ only by the position or local value of the insertion, one can build the corresponding left and right double-layer environments once and evaluate all local insertions by local contractions.
This is the double-layer analogue of the single-layer amplitude reuse described above.

The same environment idea is also used in MCMC sampling.
In a left-to-right local Metropolis sweep, one first constructs all right environments \(R_i(x)\) by a backward sweep.
Then the left environment is updated on the fly during the sweep.
At site \(i\), a trial value \(x_i'\) is tested using
\begin{equation}
W(x_1',\ldots,x_{i-1}',x_i',x_{i+1},\ldots,x_N)=L_i A_i^{x_i'}R_i(x),
\end{equation}
where \(L_i\) contains the accepted updates on the sites to the left, and \(R_i(x)\) contains the unchanged sites to the right.
After the update decision, \(L_i\) is updated to \(L_{i+1}\) using the accepted local tensor.
Therefore, the standard MC and shadow estimators benefit from the same reuse in the sampling stage, because they use the same Metropolis sampling procedure.

Regarding observable evaluation, the benefit of environment reuse depends on the Hamiltonian or on the set of observables to be evaluated.
However, for all common physical Hamiltonians considered in this work, they all give $O(N)$ reduction.
First, the reuse strategy described above applies directly to all one-body and nearest-neighbor terms.
For long-range two-body spin Hamiltonians, such as Eq.~\ref{eq:H_spin}, or four-index fermionic Hamiltonians, such as Eq.~\ref{eq:H_fermion}, one can reuse environments while sweeping over one of the site or orbital indices and explicitly looping over the remaining indices.
Thus, in all cases considered here--one-body, nearest-neighbor, long-range spin, and long-range fermionic Hamiltonians--environment reuse gives an \(O(N)\) reduction in the contraction cost.

In summary, environment reuse affects the methods considered here as shown in Table~\ref{tab:env_reuse_components}.

\begin{table}[H]
\caption{Number of needed TN contractions affected by environment reuse. 0 indicates no TN contractions needed either with or without the environment reuse.}
\label{tab:env_reuse_components}

\renewcommand{\arraystretch}{1.5}
\begin{tabular}{|c|c|c|}
\hline
Cost change & Sampling & Observable evaluation\\
\hline
Direct contraction & 0 & reduced by \(O(N)\)\\
\hline
Standard MC & reduced by \(O(N)\) & reduced by \(O(N)\)\\
\hline
Shadow estimator & reduced by \(O(N)\) & 0\\
\hline
\end{tabular}
\end{table}

For direct contraction, the relevant cost is the full observable-evaluation cost, and environment reuse reduces the cost of evaluating many locally different observables by a factor of \(O(N)\).
For standard MC, the per-sample cost consists of both sampling and local-observable evaluation; environment reuse reduces both parts by a factor of \(O(N)\).
For shadow estimators, the local-observable evaluation after sampling does not require additional tensor-network contractions, so the relevant tensor-network cost is the sampling cost; environment reuse reduces this shared sampling cost by a factor of \(O(N)\).
Thus environment reuse lowers the absolute contraction cost of all approaches considered here, while leaving their relative asymptotic comparison unchanged.

This is the standard left/right-environment reuse used in MPS and DMRG algorithms.
For PEPS and other higher-dimensional tensor networks, the same principle applies with approximate environments, such as boundary-MPS or CTMRG environments.
The details depend on the chosen approximate contraction method, but the idea is the same: store partial contractions and reuse them for amplitudes or local insertions that differ only locally.
This type of intermediate reuse is routinely used in tensor-network variational Monte Carlo and PEPS gradient calculations~\cite{PhysRevLett.99.220602,PhysRevB.83.134421,PhysRevLett.133.260404}.

\subsection{Gradient truncation}\label{app:grad_trunc}
We provide a simple motivation for the gradient-truncation approximation used in the variational ground-state calculations.
Starting from the decomposition in Eq.~\ref{eq:gradient_decomposition}, the gradient with respect to tensor parameters on site $i$ can be written as a sum of contributions from local energy terms,
\begin{equation}
\partial_{\boldsymbol{\theta}_{i}}E
=
\sum_{j}\partial_{\boldsymbol{\theta}_{i}}E_{j}.
\end{equation}
Each term has the same covariance-estimator structure as Eq.~\ref{eq:standard_gradient}, with the total local energy replaced by the local contribution $E_j$.
Although the mean contribution from a distant local term is expected to be small, its Monte Carlo estimator need not have a small variance.
This produces an artificial variance contribution from many distant pairs $(i,j)$.

This point is already visible in the non-interacting limit, where the wavefunction is a product state and the sampled configuration factorizes as
\begin{equation}
p(n)=\prod_{l}p_l(n_l).
\end{equation}
For a local energy estimator $E_j(n_j)$ supported near site $j$, the estimator for the contribution of $E_j$ to the gradient on site $i$ takes the form
\begin{equation}
g_{ij}(n)=\partial_{\boldsymbol{\theta}_{i}}\log p_i(n_i)\left(E_j(n_j)-\langle E_j\rangle\right),
\end{equation}
up to the same convention-dependent factors as in Eq.~\ref{eq:standard_gradient}.
For $i\neq j$ in the non-interacting limit, $\partial_{\boldsymbol{\theta}_{i}}\log p_i(n_i)$ and $E_j(n_j)$ are independent.
Therefore
\begin{equation}
\begin{aligned}
\langle g_{ij}\rangle
&=
\left\langle \partial_{\boldsymbol{\theta}_{i}}\log p_i(n_i)\right\rangle
\left\langle E_j-\langle E_j\rangle\right\rangle\\
&=0,
\end{aligned}
\end{equation}
because the second factor has zero mean.
However, the variance is generally nonzero,
\begin{equation}
\operatorname{Var}[g_{ij}]
=
\left\langle \left(\partial_{\boldsymbol{\theta}_{i}}\log p_i(n_i)\right)^2\right\rangle
\left\langle \left(E_j-\langle E_j\rangle\right)^2\right\rangle,
\end{equation}
provided that both the local logarithmic derivative and the local estimator fluctuate.
This argument applies to both the standard MC estimator and the shadow estimators: even when the mean contribution from a distant term vanishes, its stochastic estimator can still contribute $O(1)$ variance.

For an area-law entangled state with finite correlation length, we expect the mean contribution $\partial_{\boldsymbol{\theta}_{i}}E_j$ to decay exponentially with the distance between $i$ and $j$.
It is therefore natural to approximate the full gradient by keeping only terms within a distance cutoff,
\begin{equation}
\partial_{\boldsymbol{\theta}_{i}}E
\approx
\sum_{j:\,r_{ij}\leq R_{\mathrm{g}}}
\partial_{\boldsymbol{\theta}_{i}}E_j.
\end{equation}
This removes the stochastic variance from distant terms whose mean contribution is exponentially small.

Since long-range Hamiltonians are also considered in this work, we apply a pair decomposition  $E=\sum_{jk} E_{jk}$ in this case, and truncate the gradient as
\begin{equation}
\partial_{\boldsymbol{\theta}_{i}}E
\approx
\sum_{j,k:\,\text{min}(r_{ij},r_{ik})\leq R_{\mathrm{g}}}
\partial_{\boldsymbol{\theta}_{i}}E_{jk}.
\end{equation}

The truncation should be understood as an approximation to the exact gradient.
In particular, for standard MC the exact local-energy estimator satisfies the zero-variance principle at an eigenstate only when all Hamiltonian terms are retained.
Dropping distant terms breaks the exact cancellation responsible for the zero-variance principle.
Thus gradient truncation is useful for reducing stochastic noise, especially for the shadow estimators that do not possess the standard-MC zero-variance principle, but it is not compatible with preserving the exact zero-variance property of the full standard-MC gradient estimator.

\subsection{Bounded variance of shadow-gradient estimators} \label{app:stability_proof}
We prove the variance bound used in Sec.~\ref{subsec:stability}.
Let \(\{|x\rangle\}\) denote a discrete or continuous measurement ensemble satisfying the resolution of identity
\begin{equation}
\int d\mu(x)\,|x\rangle\langle x|=I .
\end{equation}
For a normalized parameterized state \(|\psi(\theta)\rangle\), define the Born probability
\begin{equation}
p_{\theta}(x)=|\langle x|\psi(\theta)\rangle|^{2} .
\end{equation}
We consider an estimator of the form
\begin{equation}
\begin{aligned}
E(\theta)
&=\mathbb{E}_{x\sim p_{\theta}(\cdot)}
\left[E_{\mathrm{loc}}(x)\right]\\
&=\int d\mu(x)\,
p_{\theta}(x)E_{\mathrm{loc}}(x),
\end{aligned}
\end{equation}
where \(E_{\mathrm{loc}}(x)\) is independent of \(\theta\) except through the sampled distribution \(p_{\theta}\).
This is the relevant form for the shadow estimators discussed in the main text, where \(E_{\mathrm{loc}}(x)\) denotes the bounded single-sample shadow value of a local observable contribution.
The gradient estimator is
\begin{equation}
\begin{aligned}
g_{\mathrm{loc}}(x)
&=\left(E_{\mathrm{loc}}(x)-E\right)
\partial_{\theta}\log p_{\theta}(x)
\end{aligned}
\end{equation}
It is unbiased because
\begin{equation}
\begin{aligned}
\mathbb{E}_{p_{\theta}}[g_{\mathrm{loc}}]
&=\int d\mu(x)\,
\left(E_{\mathrm{loc}}(x)-E\right)
\partial_{\theta}p_{\theta}(x)\\
&=\int d\mu(x)\,
E_{\mathrm{loc}}(x)\partial_{\theta}p_{\theta}(x)\\
&=\partial_{\theta}E(\theta),
\end{aligned}
\end{equation}
where we used \(\int d\mu(x)\,\partial_{\theta}p_{\theta}(x)=\partial_{\theta}1=0\).

The variance is bounded by the second moment,
\begin{equation}
\begin{aligned}
\operatorname{Var}\left[g_{\mathrm{loc}}(x)\right]
&\le
\mathbb{E}_{p_{\theta}}
\left[|g_{\mathrm{loc}}(x)|^{2}\right]\\
&\le
\max_x|E_{\mathrm{loc}}(x)-E|^{2}
\int d\mu(x)\,
\frac{|\partial_{\theta}p_{\theta}(x)|^{2}}{p_{\theta}(x)} .
\end{aligned}
\end{equation}
For Born probabilities,
\begin{equation}
\partial_{\theta}p_{\theta}(x)
=
2\operatorname{Re}\left[
\langle \partial_{\theta}\psi(\theta)|x\rangle
\langle x|\psi(\theta)\rangle
\right],
\end{equation}
and hence
\begin{equation}
\begin{aligned}
|\partial_{\theta}p_{\theta}(x)|^{2}
&\le
4p_{\theta}(x)
|\langle x|\partial_{\theta}\psi(\theta)\rangle|^{2} .
\end{aligned}
\end{equation}
Therefore,
\begin{equation}
\begin{aligned}
\int d\mu(x)\,
\frac{|\partial_{\theta}p_{\theta}(x)|^{2}}{p_{\theta}(x)}
&\le
4\int d\mu(x)\,
|\langle x|\partial_{\theta}\psi(\theta)\rangle|^{2}\\
&=
4\Vert\partial_{\theta}|\psi(\theta)\rangle\Vert_{2}^{2},
\end{aligned}
\end{equation}
where the last equality follows from the resolution of identity.
Combining the above inequalities gives the centered bound
\begin{equation}
\begin{aligned}
\operatorname{Var}\left[g_{\mathrm{loc}}(x)\right]
&\le
4\Vert\partial_{\theta}|\psi(\theta)\rangle\Vert_{2}^{2}
\max_x|E_{\mathrm{loc}}(x)-E|^{2}.
\end{aligned}
\end{equation}
If one instead bounds the centered local estimator by the uncentered maximum, then \(|E|\le \max_x|E_{\mathrm{loc}}(x)|\) and thus
\begin{equation}
\max_x|E_{\mathrm{loc}}(x)-E|^{2}
\le
4\max_x|E_{\mathrm{loc}}(x)|^{2}.
\end{equation}
This gives the simpler bound used in the main text,
\begin{equation}
\begin{aligned}
\operatorname{Var}\left[g_{\mathrm{loc}}(x)\right]
&\le
16\Vert\partial_{\theta}|\psi(\theta)\rangle\Vert_{2}^{2}
\max_x|E_{\mathrm{loc}}(x)|^{2}.
\end{aligned}
\end{equation}
Thus the gradient estimator has bounded variance whenever the sampled local estimator \(E_{\mathrm{loc}}(x)\) is uniformly bounded and the tangent vector \(\partial_{\theta}|\psi(\theta)\rangle\) has bounded norm.
Importantly, this conclusion does not require \(\partial_{\theta}\log p_{\theta}(x)\) to be pointwise bounded; the apparent small-probability singularity is canceled by the Born probability in the second moment.

\subsection{Absence of a comparable bound for standard Monte Carlo} \label{app:stability_counterexample}
We give a simple two-state example showing that bounded \(\|\partial_{\theta}\psi_{\theta}\|\) does not imply a bounded variance for the standard Monte Carlo energy-gradient estimator.
Consider
\begin{equation}
\begin{aligned}
|\psi_{\epsilon}\rangle
&=
\epsilon |0\rangle
+
\sqrt{1-\epsilon^{2}}|1\rangle,\\
H&=|0\rangle\langle1|+|1\rangle\langle0|,
\end{aligned}
\end{equation}
with parameter \(\theta=\epsilon\).
Then
\begin{equation}
\partial_{\epsilon}|\psi_{\epsilon}\rangle
=
|0\rangle
-
\frac{\epsilon}{\sqrt{1-\epsilon^{2}}}|1\rangle,
\end{equation}
so \(\|\partial_{\epsilon}\psi_{\epsilon}\|\) remains bounded as \(\epsilon\to0\).
The Born probabilities are
\begin{equation}
p_{\epsilon}(0)=\epsilon^{2},
\qquad
p_{\epsilon}(1)=1-\epsilon^{2} .
\end{equation}
At configuration \(x=0\), the local energy is
\begin{equation}
\begin{aligned}
E_{\mathrm{loc}}(0)
&=
\frac{\langle0|H|\psi_{\epsilon}\rangle}
     {\langle0|\psi_{\epsilon}\rangle}\\
&=
\frac{\sqrt{1-\epsilon^{2}}}{\epsilon}
\sim
\frac{1}{\epsilon} .
\end{aligned}
\end{equation}
The local energy is singular, but this singularity alone does not make the standard Monte Carlo energy estimator unstable.
Indeed, at the other configuration,
\begin{equation}
\begin{aligned}
E_{\mathrm{loc}}(1)
&=
\frac{\langle1|H|\psi_{\epsilon}\rangle}
     {\langle1|\psi_{\epsilon}\rangle}\\
&=
\frac{\epsilon}{\sqrt{1-\epsilon^{2}}}.
\end{aligned}
\end{equation}
Thus the second moment of the energy estimator is
\begin{equation}
\begin{aligned}
\mathbb{E}_{p_{\epsilon}}\left[|E_{\mathrm{loc}}|^{2}\right]
&=
p_{\epsilon}(0)|E_{\mathrm{loc}}(0)|^{2}
+p_{\epsilon}(1)|E_{\mathrm{loc}}(1)|^{2}\\
&=
\epsilon^{2}\frac{1-\epsilon^{2}}{\epsilon^{2}}
+(1-\epsilon^{2})\frac{\epsilon^{2}}{1-\epsilon^{2}}\\
&=
1 .
\end{aligned}
\end{equation}
Therefore the ordinary energy estimator has bounded variance in this example.
The problem appears only after multiplying the local energy by the score factor in the gradient estimator.
This factor satisfies
\begin{equation}
\partial_{\epsilon}\log p_{\epsilon}(0)
=
\frac{2}{\epsilon} .
\end{equation}
Thus the one-sample gradient contribution at \(x=0\) scales as
\begin{equation}
\begin{aligned}
E_{\mathrm{loc}}(0)
\partial_{\epsilon}\log p_{\epsilon}(0)
&\sim
\frac{2}{\epsilon^{2}} .
\end{aligned}
\end{equation}
Its contribution to the second moment is therefore
\begin{equation}
\begin{aligned}
&p_{\epsilon}(0)
\left|
E_{\mathrm{loc}}(0)
\partial_{\epsilon}\log p_{\epsilon}(0)
\right|^{2}\\
&\qquad\sim
\epsilon^{2}\frac{4}{\epsilon^{4}}
=
\frac{4}{\epsilon^{2}},
\end{aligned}
\end{equation}
which diverges as \(\epsilon\to0\).
Hence the standard Monte Carlo gradient estimator has no uniform variance bound of a form controlled only by \(\|\partial_{\theta}\psi_{\theta}\|\), even though the corresponding energy estimator remains well behaved.

\end{document}